\documentclass[aps,prc,twocolumn,amssymb,amsmath,superscriptaddress,floatfix,showpacs,nofootinbib]{revtex4}
\usepackage[dvips]{graphics,graphicx}
\usepackage{multirow}
\usepackage{longtable}
\begin{document}
\title{Critical behavior of charmonia across the phase transition: A QCD sum
rule approach}
\author{Kenji Morita}
\email{morita@phya.yonsei.ac.kr}
\author{Su Houng Lee}
\email{suhoung@phya.yonsei.ac.kr}
\affiliation{Institute of Physics and Applied Physics, Yonsei
University, Seoul 120-749, Korea}
\date{\today}
\begin{abstract}
 We investigate medium-induced change of mass and width of $J/\psi$ and
 $\eta_c$ across the phase transition in hot gluonic matter using QCD
 sum rules. In the QCD sum rule approach,
 the medium effect on heavy quarkonia is induced by the change of
 both scalar and twist-2 gluon condensates, whose temperature dependences
 are extracted from the lattice calculations of energy density and
 pressure.
 Although the stability of the operator product expansion side
 seems to break down at
 $T>1.06T_{\text{c}}$ for the vector channel and $T>1.04T_{\text{c}}$
 for the pseudoscalar
 channel, we find a sudden change of the spectral property across the
 critical temperature $T_{\text{c}}$, which originates from an equally
 rapid change of the scalar gluon condensate characterized by
 $\varepsilon -3p$. By parameterizing the
 ground state of the spectral density by the Breit-Wigner form,
 we find that for  both $J/\psi$ and
 $\eta_c$, the masses suddenly decrease maximally by a few hundreds of
 MeV and the widths broaden to $\sim$ 100 MeV slightly above $T_{\text{c}}$.
 Implications for recent and future heavy ion experiments are discussed.
 We also carry out a similar analysis for charmonia in nuclear matter,
 which could serve as a testing ground for observing the precursor
 phenomena of the QCD phase transition. We finally discuss the
 possibility of observing the mass shift at nuclear matter at the FAIR
 project at GSI.
\end{abstract}

\maketitle

\section{Introduction}

In-medium change of spectral properties of heavy quarkonia is one of the
interesting problems in recent hadron physics. Firstly, the recent
relativistic heavy ion collision experiment at the Relativistic Heavy
Ion Collider (RHIC) reveals exciting nature of the QCD matter through a
number of observations
\cite{BRAHMS_Whitepaper,PHOBOS_Whitepaper,STAR_Whitepaper,PHENIX_Whitepaper,Gyulassy05}.
However, there are many open questions in
both experimental facts and theoretical understandings of QCD
matter. Hence, it is important
to establish appropriate experimental observables that reflect consequences of
deeper theoretical understanding of the matter. Heavy
quarkonia have been regarded as one of the most suitable diagnostic
tools in this respect, since the suppression of
$J/\psi$ yields would reflect the Debye screening phenomenon caused by
the deconfinement phenomenon in the quark-gluon
plasma (QGP), as was originally argued by Matsui and Satz
\cite{Matsui_PLB178}. Until now, quarkonium production,
especially that of $J/\psi$, in relativistic heavy ion collisions have
been extensively studied both experimentally
\cite{PHENIX_Jpsi04,PHENIX_Jpsilatest} and theoretically
\cite{Kharzeev_QM06,andronic07_plb652}. However, a remarkable progress comes from recent lattice
QCD calculations, which indicate that contrary to the earlier
expectation the $J/\psi$ will survive as a bound state even in the QGP up to
$T\sim 1.6-2T_{\text{c}}$
\cite{Umeda_IJMP16,Asakawa_PRL92,Datta_PRD69}, which was anticipated
before based on the non-perturbative nature of QGP
\cite{hansson88}. Nowadays, the state of
matter at this temperature region has been characterized as ``strongly
coupled'' QGP (sQGP).
Hence, there will be change of spectral properties even for heavy quark
system which has to be considered in interpreting experimental observables.

Secondly, charmonium in a nuclear medium is also an interesting issue.
In relativistic heavy ion collisions, we need knowledge of
quarkonium-nucleon interaction to discriminate the suppression by QGP
from the ``cold nuclear matter effect'' induced by such an interaction.
Furthermore, multi-gluon exchange can lead to an attractive interaction
between $c\bar{c}$ and a nucleon, which may result in a bound state of
charmonium and a light nuclei, as pointed out by Brodsky \textit{et
al.}\cite{Brodsky90}. It should be also noted that the Panda
experiment at GSI-FAIR plans reaction of anti-protons with nucleus
target, which will yield charmonia in the nuclear matter. It could serve
as a testing ground for observing the precursor phenomenon of the
QCD phase transition.

In this paper, we investigate change of mass and width of $J/\psi$ and
$\eta_c$ induced by strongly interacting hot gluonic matter and by nuclear
medium using QCD sum rule. QCD sum
rule provides a systematic procedure for studying hadrons from a
viewpoint of the asymptotic freedom in
QCD \cite{Shifman_NPB147,Shifman_NPB147_2}. Since the QCD sum rule
can take non-perturbative effects into account through the condensate
terms, it is a suitable
theoretical tool of
the current study. Indeed, QGP at $T < 3T_{\text{c}}$ cannot be
understood using perturbation theory alone \cite{Blaizot_QM06}. Furthermore, the sum rule is
more promising for heavy quark systems because we do not have to
take the quark-antiquark condensate into account unlike light quark
systems. In this respect, the sum rule has been applied to
charmonium and bottonium. Shifman \textit{et al.} established the
framework in Ref.~\cite{Shifman_NPB147,Shifman_NPB147_2} and
Reinders \textit{et al.}
extended it to deep Euclidean region $Q^2 = -q^2 > 0$
~\cite{Reinders_NPB186}, in the case of vacuum. As for the quarkonia
in-medium, One of us together with Furnstahl and Hatsuda have
investigated the mass shift of
$J/\psi$ in hot hadronic matter \cite{Furnstahl_PRD42}, using a QCD sum
rule approach, where the temperature effect was introduced to the
perturbative Wilson coefficient through the scattering terms. A
consistent formalism at lower density was developed by one of us
\cite{Klingl_PRL82} and independently by Hayashigaki
\cite{Hayashigaki99:_jpsi} to study the mass shift of $J/\psi$ in
nuclear matter.

Along this direction, we investigated the mass shift and width broadening
of $J/\psi$ in hot gluonic plasma (GP) \cite{Morita_jpsiprl} just
above the phase transition by consistently using the exact temperature
dependencies of condensates from lattice calculation. In the
present paper, as a
subsequent paper to Ref.~\cite{Morita_jpsiprl}, we present details of
the analysis, further application to $\eta_c$ and to spectral changes in
nuclear matter.

In the next section, we will give an explanation of the QCD sum rule for heavy
quarkonium in medium used in the present work. Section \ref{sec:GP} and
\ref{sec:NM} describe the details of the numerical computations of the sum
rule for hot gluonic matter and nuclear medium, respectively. Section
\ref{sec:summary} is devoted to discussion and summary.

\section{QCD sum rule for heavy quarkonium}

In this section, first we review the sum rule for heavy quarkonium in
vacuum \cite{Reinders_NPB186}. Then we introduce the extension to finite
temperature and nuclear medium cases, in which medium effect is
eventually induced only by the expectation values of gluonic operators
without any additional change in the operator product expansion (OPE).

\subsection{Moment sum rule in vacuum}

We start with the time-ordered current-current correlation function for
$J$ channel
\begin{equation}
 \Pi^{J}(q) = i \int d^4 x \, e^{iq\cdot x}
  \langle T[j^{J}(x)j^{J}(0)]\rangle,\label{eq:correlation}
\end{equation}
where we consider $J=P$ (pseudoscalar) and $V$ (vector) current of the
heavy quark. Namely,
$j^P = i \bar{c}\gamma_5 c$ and $j^{V}_\mu = \bar{c}\gamma_\mu c$, for
charm. The expectation value $\langle \cdots \rangle$ is taken for
the vacuum. If we go to deep Euclidean
region $Q^2 \equiv -q^2  \gg 0$,
the product of the current can be expanded via operator production
expansion (OPE) \cite{Wilson69}. If we denote $\tilde{\Pi}(q^2)$ such
that $\Pi^{\mu\nu}(q)=(q^\mu q^\nu-q^2 g^{\mu\nu})\tilde{\Pi}(q^2)$ for
the vector current, $\tilde{\Pi}(q^2)$ can be written as
\begin{equation}
 \tilde{\Pi}^J(q^2) = \sum_{n}C_n^J \langle O_n \rangle\label{eq:Pitilde}
\end{equation}
where $O_n$ are the operators of mass dimension $n$ renormalized at
scale $\mu^2$ and $C_n^J$ are the Wilson coefficient.
By virtue of much heavier quark mass than the confinement scale, heavy quark
operators, such as $m_c\bar{c}c$
for dimension 4, are rewritten in terms of gluonic operator with a
factor of $1/m_c$ via heavy
quark expansion \cite{Shifman_NPB147,generalis84}. Hence, only gluonic
operators contribute to the OPE for the heavy quark currents.

On the other hand, the correlation function \eqref{eq:Pitilde} is
related to its imaginary part through the dispersion relation
\begin{equation}
 \tilde{\Pi}^J(q^2) = \frac{1}{\pi}\int_{4m_c^2}^{\infty} \frac{\text{Im}\tilde{\Pi}^J(s)}{s-q^2}ds\label{eq:disp1}
\end{equation}
where we ignore $+i\varepsilon$ in the denominator of the integrand
since $q^2 =-Q^2 < 0$.
Taking $n$ times derivative of Eqs.~\eqref{eq:Pitilde} and
\eqref{eq:disp1} as
\begin{equation}
 M_n^J(Q^2) \equiv \left.\frac{1}{n!}\left(\frac{d}{dq^2}\right)^n
		  \tilde{\Pi}^J(q^2)\right|_{q^2 = -Q^2},
\end{equation}
we obtain the $n$-th order moment for the OPE side
\begin{equation}
 M_n^J(Q^2)_{\text{OPE}}=  A_n^J(\xi)[1+a_n^J(\xi)\alpha_{\text{s}}
 +b_n^J(\xi)\phi_{\text{b}}],\label{eq:moment_ope}
\end{equation}
and that for the phenomenological (dispersion) side
\begin{equation}
 M_n^J(Q^2)_{\text{phen.}} = \frac{1}{\pi}\int_{4m_c^2}^{\infty}
 \frac{\text{Im}\tilde{\Pi}^J(s)}{(s+Q^2)^{n+1}} ds .\label{eq:dispint}
\end{equation}
Here, we have introduced a dimensionless scale variable $\xi = Q^2/4m_c^2$.
In Eq.~\eqref{eq:moment_ope}, $A_n^J(\xi)$, $a_n^J(\xi)$, and
$b_n^J(\xi)$ are the Wilson coefficients which correspond to bare loop
diagrams, perturbative radiative correction up to order $\alpha_s$, and
scalar gluon condensate, respectively.
These coefficients were derived in Ref.~\cite{Reinders_NPB186}
and we summarize them in the Appendix.

In evaluation of spectral properties, we take the ratio of the ($n-1$)-th
moment to the $n$-th moment and equate the OPE side with the
phenomenological side. Then we obtain the sum rule
\begin{equation}
 \left.\frac{M_{n-1}^J}{M_n^J}\right|_{\text{OPE}} = \left.\frac{M_{n-1}^J}{M_n^J}\right|_{\text{phen.}},\label{eq:sumrule}
\end{equation}
which relates the hadron properties (r.h.s.) with asymptotically free QCD
(l.h.s.)

\subsection{Moment sum rule for the hot gluonic medium}

In this paper, we firstly consider the gluonic medium at finite
temperature around $T_{\text{c}}$. Then, the expectation value in
Eq.~\eqref{eq:correlation} is taken as
$\langle \mathcal{O} \rangle = \text{Tr}(e^{-\beta H} \mathcal{O})/\text{Tr}
(e^{-\beta H})$. Hereafter, we set both medium and $c\bar{c}$ at rest.
We denote $q^\mu = (\omega,\boldsymbol{q})$ and take
$\boldsymbol{q}\rightarrow 0$ limit. In this case, the transverse and
the longitudinal components of the correlation function for the vector
channel are simply related with
$\Pi_{\text{T}} = \omega^2\Pi_{\text{L}}$
and
$\Pi_{\text{L}} = \Pi_\mu^\mu / (-3\omega^2)$.
We denote the longitudinal component as $\tilde{\Pi}^J(\omega)$ for the
vector channel.

At finite temperature, \textit{retarded} correlation function is related to the
spectral function \cite{Lifshitz80:_statis_physic_part}. In the
Euclidean region $\omega^2 < 0$, the retarded correlation function
$\Pi^R(\omega)$ becomes $\Pi(\omega^2)$ and the dispersion
relation is given by \cite{Furnstahl_PRD42,Hatsuda93}
\begin{equation}
 \tilde{\Pi^J}(\omega^2) = \int_{0^-}^{\infty}du^2
  \frac{\rho(u)}{u^2-\omega^2},\label{eq:dist_atfiniteT}
\end{equation}
where $\rho(u)$ is the spectral function connecting with the imaginary
part as
\begin{equation}
 \rho(u) =
  \frac{1}{\pi}\tanh\left(\frac{u}{2T}\right)\text{Im}\tilde{\Pi}^J(u^2).
\end{equation}
Then Eq.~\eqref{eq:dist_atfiniteT} reduces to the vacuum case
[Eq.~\eqref{eq:disp1}] when $\text{Im}\tilde{\Pi}^J(u^2)$ has nonzero
value only at $u \gg T$. Since we are interested in charmonia for which
the mass is much larger than temperature considered here, this condition
seems to be appropriate one. However, there are formally two additional
terms in the finite temperature spectral function
\cite{Bochekarev86}. One is the continuum part which also exists in the
case of vacuum. Following the prescription in
Ref.~\cite{Reinders_NPB186}, we can suppress contribution from this part
as described later because this part has finite values beyond some
threshold. The other part arising from scattering of the current
with quarks in medium is proportional to $\delta(u^2)$ and the
contribution grows
up with $T$ in the hadronic medium \cite{Furnstahl_PRD42}.
However, since we are considering the \textit{gluonic} medium
in which there are no (anti-)quarks which annihilate with the current,
such a scattering term does not appear.
Hence, we can use the same expression of the phenomenological side with
the vacuum case [Eq.~\eqref{eq:dispint}] for charmonia in the hot
gluonic medium.

As for the OPE side, there is an important change from the vacuum to the
medium case. Since we have no longer Lorentz invariance, non-scalar
operators have non-vanishing value \cite{Hatsuda93}. In the present
case, twist-2 gluon operator has leading contribution and the $n$-th
order moment of the OPE side [Eq.~\eqref{eq:moment_ope}] should be
modified to
\begin{equation}
 M_n^J(Q^2)_{\text{OPE}}=  A_n^J(\xi)[1+a_n^J(\xi)\alpha_{\text{s}}
 +b_n^J(\xi)\phi_{\text{b}}+c_n^J(\xi)\phi_{\text{c}}],
 \label{eq:moment_ope_medium}
\end{equation}
where $c_n$ and $\phi_{\text{c}}$ are the Wilson coefficients and the
medium expectation value for the twist-2 operator. Since we are
considering the heavy quark systems, only the condensate terms are
temperature dependent as long as $T \ll m_c, |Q|$
\cite{Furnstahl_PRD42,Hatsuda93}. Hence, the Wilson coefficients are the
same as in the vacuum case.
In the following, we
show that the gluon condensates $\phi_{\text{b,c}}$ are written in
terms of thermodynamic quantities which can be extracted from lattice
QCD data.

\begin{figure}[ht!]
 \includegraphics[width=3.375in]{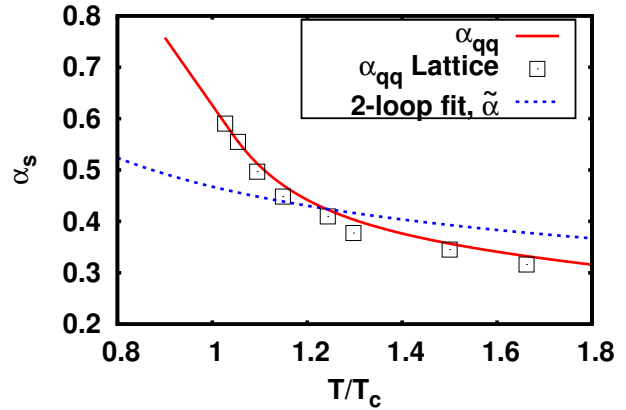}
 \caption{(Color online) Temperature dependent coupling constants extracted from lattice
 QCD. The boxes denote the lattice data points of
 $\alpha_{\text{qq}}(r_{\text{screen}},T)$ taken from
 Ref.~\cite{Kaczmarek_PRD70}. The solid line is drawn by Bezier
 interpolation of the lattice data points. The dotted line shows the
 case of Eq.~\eqref{eq:gpert}.}
 \label{fig:alpha}
\end{figure}

If we define these condensate terms as
\begin{align}
 G_0(T) &= \left\langle
 \frac{\alpha_{\text{s}}}{\pi}G^a_{\mu\nu}G^{a\mu\nu}\right\rangle_T,\label{eq:defG0}\\
 \left(u^\mu u^\nu - \frac{1}{4}g^{\mu\nu} \right)G_2(T) &= \left\langle
 \frac{\alpha_{\text{s}}}{\pi}G^{a\mu}_\rho
 G^{a\nu\rho}\right\rangle_T,\label{eq:defg2}
\end{align}
where $u^\mu$ is the 4-velocity of the medium and taken to be
$u^\mu=(1,0,0,0)$, explicit forms of $\phi_{\text{b,c}}$ are given as
\begin{align}
 \phi_{\text{b}}&=\frac{4\pi^2}{9(4m_c^2)^2}G_0(T),\label{eq:phib_qgp}\\
 \phi_{\text{c}}&=\frac{4\pi^2}{3(4m_c^2)^2}G_2(T).\label{eq:phic_qgp}
\end{align}
Actually it is possible to calculate the condensates \eqref{eq:defG0}
and \eqref{eq:defg2} directly using
lattice QCD, but we do not adopt such an approach here. The gluon condensates
generally consist of the perturbative piece and the non-perturbative
piece. At zero temperature, the condensate term appearing in QCD sum
rules is the non-perturbative piece only and it is shown that the
non-perturbative part extracted from lattice QCD by subtracting the
perturbative part is indeed consistent with the value of the condensate
determined from QCD sum rules for charmonium
\cite{Giacomo81,Giacomo82,Campostrini87}. Similar consideration holds
also for the finite temperature case \cite{Lee89}, in which we would
have to subtract
out the perturbative part at $T \neq 0$ if we directly calculated the
non-perturbative condensates from lattice QCD. In this paper, since we
are putting all the temperature dependencies in the condensates,
including the perturbative and non-perturbative contributions, we can
just extract total temperature dependencies of the operators from the
lattice. This is possible by noting that the scalar gluon condensate and
twist-2 gluon condensates are respectively just the trace part and
symmetric traceless part of the energy momentum tensor. This energy
momentum tensor is well calculated on the lattice from the pressure and
energy density of the plasma through the following equation,
\begin{equation}
 T^{\alpha \beta} = (\varepsilon+p)\left(u^\alpha u^\beta -
				    \frac{1}{4}g^{\alpha \beta}\right)
 + \frac{1}{4}(\varepsilon-3p)g^{\alpha\beta}.\label{eq:e-m-t}
\end{equation}

The scalar condensate can be related to the trace part through the trace
anomaly as
\begin{equation}
 T^{\mu}_\mu = \left\langle \frac{\beta(g)}{2g}G^{a}_{\mu\nu}G^{a\mu\nu}
				   \right\rangle,
\end{equation}
with $\beta(g)$ being the beta function,
$\beta(g)=-\frac{g^3}{48\pi^2}\left(33-2 N_f \right)$ for 1-loop,
$N_{\text{f}}$ flavors, and $N_c$ colors. Using the above expression with
$N_{\text{f}}=0$ and $N_{\text{c}}=3$ for the beta function and
recalling that $T^\mu_\mu=\varepsilon-3p$, we obtain
\begin{equation}
 G_0(T) = G_0^{\text{vac}}-\frac{8}{11}(\varepsilon-3p)\label{eq:g0}
\end{equation}
where $G_0^{\text{vac}}$ is the value of the scalar gluon condensate in
vacuum \cite{Miller07}. As for the twist-2 part, the symmetric traceless
part of the energy-momentum tensor is the gluon operator
\begin{equation}
  T^{\alpha\beta} = -G^{a\alpha\lambda}G^{a\beta}_{\lambda}.\label{eq:emtensor_gluon}
\end{equation}
Hence we can identify the traceless part of the energy momentum tensor to
$(\varepsilon+p)$ as given in Eq.\eqref{eq:e-m-t}.
>From Eq.~\eqref{eq:defg2}, the twist-2 part becomes
\begin{equation}
 G_2(T) = -\frac{\alpha_\text{s}(T)}{\pi}(\varepsilon+p),
\end{equation}
so that $G_2(T)$ is proportional to the entropy density of the system
$s=(\varepsilon+p)/T$.
We extract the temperature dependent quantities $\varepsilon$, $p$
\cite{Boyd_NPB469} and $\alpha_{\text{s}}(T)$ \cite{Kaczmarek_PRD70}
from lattice calculations for the pure SU(3) system.
In order to construct $G_2$, we need the temperature dependent effective
coupling constant. The coupling constant, however, cannot be uniquely
determined by lattice QCD
\cite{Kaczmarek_PRD70}. Ref.~\cite{Kaczmarek_PRD70} presented four kinds
of the coupling constant extracted from the color singlet heavy
quark-antiquark free energy. Two of them are measured in the short
distant regime
and the others are done in the long distant regime. In the former,
one is from the free energy and the other is from the spatial derivative of the
free energy (force).
Both coupling constants are almost independent of temperature at short
distance, $r < 0.1$ fm. While the former goes to negative value at
larger distance due to the remnant of the confinement force, the latter
shows temperature dependent maximum value, at which the distance is
denoted by $r_{\text{screen}}$. Here, we adopt the latter one,
$\alpha_{\text{qq}}(r,T)$ at $r=r_{\text{screen}}$ as one of reasonable
coupling constants since it characterizes the relevant length scale for
the separation of short distance regime from long distance one.
On the other hand, the long distant regime is based on a fit of the free
energy to the Debye-screened functional form which has two coupling
parameters, Coulomb
force strength $\alpha(T)$ and screening $\tilde{\alpha}(T)$. Although
both of the coupling constants show reasonable temperature dependencies
and agree each other at $T > 6T_{\text{c}}$, we adopt
$\tilde{\alpha}(T)$ because the Coulomb force strength is not relevant
for characterizing the long distance non-perturbative physics at temperature
considered here.
Unlike $\alpha_{\text{qq}}$, the
uncertainty in the result of $\tilde{\alpha}(T)$ is too
large. Therefore, we use
the 2-loop perturbative running coupling form
\begin{equation}
 g^{-2}_{\text{pert}}(T) =
  \frac{11}{8\pi^2}\ln\left(\frac{2\pi
		       T}{\Lambda_{\overline{\text{MS}}}}
		      \right)
  +\frac{51}{88\pi^2}\ln
  \left[2\ln\left(\frac{2\pi
	     T}{\Lambda_{\overline{\text{MS}}}}\right)\right],\label{eq:gpert}
\end{equation}
with $T_{\text{c}}/\Lambda_{\overline{\text{MS}}}\simeq 1.14$ and
rescale this as $\tilde{\alpha}(T)=2.095\alpha_{\text{pert}}(T)$
\cite{Kaczmarek_PRD70}. Here we put $T_{\text{c}}=264$ MeV
\cite{Boyd_NPB469}. The two coupling constants as a function of
temperature are displayed in Fig.~\ref{fig:alpha}. As explained later,
we will consider only temperature region near $T_{\text{c}}$ in this
paper. Hence, $\alpha_{\text{qq}}$ is stronger than $\tilde{\alpha}(T)$
throughout analyses in this paper.

\begin{figure}[ht!]
 \includegraphics[width=3.375in]{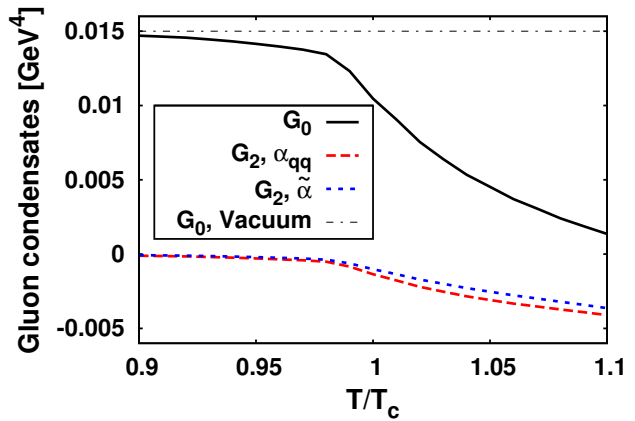}
 \caption{(Color online) Gluon condensates near $T_{\text{c}}$.}
 \label{fig:gc}
\end{figure}

The resultant gluon condensates $G_0$ and $G_2$ for two cases of the
coupling constant are shown in Fig.~\ref{fig:gc}\footnote{We have
renewed the extraction from lattice data by improving the resolution, so
that the present values are slightly different from those of
Ref.~\cite{Morita_jpsiprl}.}. For $G_0$, we use
$G_0^{\text{vac}}=(0.35 \text{GeV}{})^4 \simeq 0.015 \text{GeV}^4$. We
can see that $G_0$ decreases as temperature increases and reaches
less than half of the vacuum value at $T/T_{\text{c}}\simeq 1.04$. It
becomes negative at higher temperature but remains positive in the
temperature region considered here \cite{Lee89}.

\subsection{Moment sum rule for the nuclear medium}

In this case, the medium consist of nucleons, thus we do not have to
worry about the scattering term. As far as we follow the same method to
suppress the contribution from the continuum, we can use the same form of
the phenomenological side with the vacuum and finite temperature cases.

\begin{figure}[!ht]
 \includegraphics[width=3.375in]{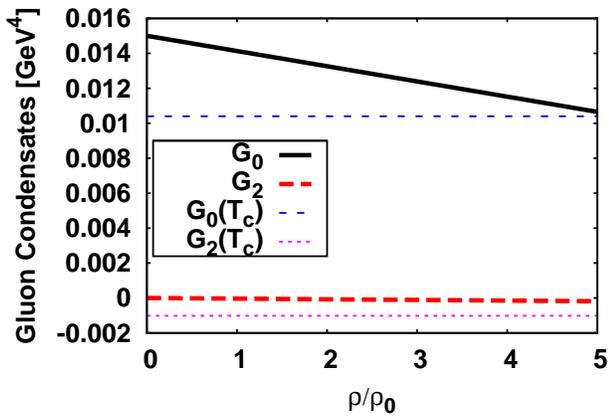}
 \caption{(Color online) Gluon condensates in nuclear matter. Thick
 solid and dashed
 line show the scalar and twist-2 condensates as a function of density
 normalized by the normal nuclear density. Thin lines are the finite
 temperature case at $T=T_{\text{c}}$ for a comparison.}
 \label{fig:gcinnm}
\end{figure}

Thus, since the medium effect is similarly imposed on the gluon condensates,
difference in the nuclear matter case from the case of hot gluonic
matter is in the explicit form of
$\phi_{\text{b,c}}$. In order to evaluate the expectation value for the
ground state of the nuclear matter, we employ the linear density
approximation \cite{Cohen92}:
\begin{equation}
 \langle O \rangle_{\text{n.m.}}=\langle O \rangle_0+\frac{\rho_N}{2m_N}
  \langle N | O | N \rangle,
\end{equation}
where $\rho_N$ and $m_N$ are the normal nuclear matter density and the
nucleon mass,
respectively. The nucleon state $|N \rangle$ is normalized as
$\langle N(p') | N(p) \rangle = 2p_0 (2\pi)^3 \delta^3 (p-p')$.
Then, the scalar condensate becomes \cite{Klingl_PRL82}
\begin{equation}
 \left\langle \frac{\alpha_{\text{s}}}{\pi}G^a_{\mu\nu}G^{a\mu\nu}
 \right\rangle_{\text{n.m.}} = \left\langle
 \frac{\alpha_{\text{s}}}{\pi}G^a_{\mu\nu}G^{a\mu\nu}\right\rangle_0
 -\frac{8}{9}m_N^0 \rho_N\label{eq:g0_nm}
\end{equation}
where $m_N^0\simeq 750$ MeV is the nucleon mass in the chiral limit
\cite{Borasoy96}.
The traceless and symmetric twist-2 operator is given as
\cite{Klingl_PRL82},
\begin{equation}
 \left\langle N(p) \left|
  \frac{\alpha_{\text{s}}}{\pi}G^a_{\alpha\sigma}G^{a\beta\sigma}
   \right| N(p) \right\rangle =
 -\left(p_\alpha p^\beta -\frac{1}{4}g_\alpha^\beta p^2
  \right)\frac{\alpha_{\text{s}}}{\pi}A_{\text{G}}
 \end{equation}
where $A_{\text{G}}$ is related to the moment of the gluon distribution
function
\begin{equation}
 A_{\text{G}}(\mu^2) = 2\int_{0}^{1}dx\, x G(x,\mu^2).
\end{equation}
Following Ref.~\cite{Klingl_PRL82}, we take
$A_{\text{G}}(8m_c^2)\simeq 0.9$. While $G_2$ at finite temperature is
related to the entropy, this correspondence does not hold in the nuclear
matter case. Note that Eq.~\eqref{eq:emtensor_gluon} does not contain the
quark sector.
Using these expressions, the condensate terms which appear in
Eq.~\eqref{eq:moment_ope} finally result in \cite{Klingl_PRL82}
\begin{align}
 \phi_{\text{b}}&=
 \frac{4\pi^2}{9(4m_c^2)^2}
 \left\langle\frac{\alpha_{\text{s}}}{\pi}G^a_{\mu\nu}G^{a\mu\nu}
 \right\rangle_{\text{n.m.}}\label{eq:gc_nm} \\
 \phi_{\text{c}}&=
 -\frac{2\pi^2}{3}\frac{\frac{\alpha_{\text{s}}}{\pi}A_{\text{G}}}{(4m_c^2)^2}
 m_N \rho_N.\label{gc2_nm}
\end{align}
The form of $\phi_{\text{b}}$ is the same as the hot gluonic matter case
but now the
expectation value is taken through Eq.~\eqref{eq:g0_nm}. We depict the
density dependence of the gluon condensates based on
Eqs.~\eqref{eq:g0_nm} and \eqref{gc2_nm} in Fig.~\ref{fig:gcinnm}.
The twist-2 case is re-normalized so that it corresponds to the finite
temperature case \eqref{eq:phic_qgp}. We can see that the change of the scalar
condensate reaches as large as $T=T_{\text{c}}$ case at $\rho \sim 5\rho_0$
but is much smaller at the normal nuclear density.
The twist-2 contribution is much smaller than that of the finite
temperature case.

\subsection{Phenomenological side}

In the phenomenological side, we use a simple prescription which describes the
lowest lying resonance in each channel. For charmonium, previous studies
\cite{Shifman_NPB147,Reinders_NPB186,Furnstahl_PRD42,Klingl_PRL82,Hayashigaki99:_jpsi}
focused on the mass and ignored the small but finite width of $J/\psi$
and $\eta_{\text{c}}$. In this case, the imaginary part of the
polarization function in Eq.~\eqref{eq:dispint} is simply parametrized
by
\begin{equation}
 \text{Im}\tilde{\Pi}(s) = f_0 \delta(s-m^2)+\text{corrections},\label{eq:phen_massless}
\end{equation}
where we ignore the channel subscript $J$. This spectral function
immediately leads to the moment
\begin{equation}
 M_n(\xi) = \frac{f_0}{\pi (m^2+Q^2)^{n+1}}[1+\delta_n(\xi)].
\end{equation}
The correction term in Eq.~\eqref{eq:phen_massless} is absorbed in
$\delta_n(\xi)$. By taking the ratio as in Eq.~\eqref{eq:sumrule}, we
can remove the constant $f_0$ from the equation. To obtain the mass of
lowest lying resonance, we need to choose sufficiently large $n$ such
that $(1+\delta_{n-1}(\xi))/(1+\delta_n(\xi))$ is close to unity. Then
the ratio does not depend on the details of the correction term which
contains higher resonances and continuum, and the mass is simply given by
\begin{equation}
 m^2 \simeq \frac{M_{n-1}(\xi)}{M_n(\xi)}-4m_c^2\xi.\label{eq:zerowidthlimit}
\end{equation}
Previous analyses rely on this formula.

In this work, we extend the above formulation to include finite width.
Here, we employ the simple relativistic Breit-Wigner form
\begin{equation}
 \text{Im}\tilde{\Pi}(s) = \frac{f_0 \sqrt{s} \Gamma}{(s-m^2)^2+s\Gamma^2}
 +\text{corrections}.\label{eq:BW}
\end{equation}
As in the $\Gamma=0$ case, we can eliminate the unnecessary
constant and the effects of the correction term by taking the ratio of the
moment and choosing appropriately large $n$. In the practical analyses
of the sum rule, our task is to find values of $(m,\Gamma)$ which
satisfy the sum rule [Eq.~\eqref{eq:sumrule}]. Generally there are infinite
numbers of the pairs of $(m,\Gamma)$ because the sum rule provides one
equation with respect to the two quantities which we want to know.
Hence, without additional constraints, the sum rule can provide only
relation between $m$ and $\Gamma$ as in the case of light vector mesons
\cite{Leupold98}. Here, before the practical calculation, we discuss the
relation between the mass and the width which comes from the
phenomenological side, Eq.~\eqref{eq:BW}.

In calculation of the moment ratio of the phenomenological side, we need
to compute the dispersion integral in Eq.~\eqref{eq:dispint} with the spectral
function in Eq.~\eqref{eq:BW}. Since the width of the ground state charmonium
is much smaller than its mass, we need careful treatment in numerical
integration. To achieve good accuracy, we performed Monte-Carlo
integration based on the VEGAS algorithm
\cite{Press96:_numer_recip_fortr}. In our calculation, typical
relative numerical uncertainty evaluated from the standard manner in the
Monte-Carlo integration is order of $10^{-6}$ for $10^6$
events with $m=3$ GeV and $\Gamma=1$ MeV. As expected, this accuracy
becomes better as $\Gamma$ increases.

\begin{figure}[ht!]
 \includegraphics[width=3.375in]{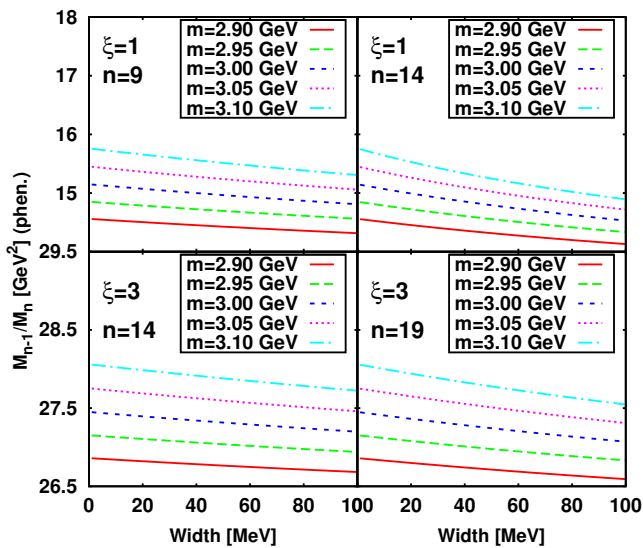}
 \caption{(Color online) Moment ratio of the phenomenological side as a
 function of
 $\Gamma$. Upper panels are for $\xi=1$. Left and right panels denote the
 case of $n=9$ and $n=14$, respectively. Lower ones are for $\xi=3$ with
 $n=14$ (left) and $n=19$ (right).}
 \label{fig:phen}
\end{figure}

We plot the $\Gamma$ dependence of the moment ratio for various mass values
from 2.9 GeV to 3.1GeV in Fig.~\ref{fig:phen}. Here we show the result
for two values of $\xi$, $\xi=1$ and
3. In each $\xi$ value, we choose two values of $n$, of which are the
typical values for the analyses below, to see $n$ dependence of the
moment ratio. First, comparing the left (smaller $n$) side to the right
(large $n$) side, we can see that $\Gamma$ dependence of the moment
ratio becomes stronger as $n$ increases. As we will see later,
larger $n$ is suitable for evaluating mass at higher temperature. Hence,
this fact means that, as the temperature increase, the
system becomes more sensitive to the width. Second, the moment ratio
decreases monotonically as the width increases if mass is unchanged. It
also decreases as the
mass decreases but the width dependence is much weaker. For instance,
let us suppose that we obtain 1 GeV$^2$ decrease of the moment ratio
from the OPE side for $\xi=1$. If mass stays constant, the width must broaden
to larger than 100 MeV while it corresponds to about 100 MeV mass
reduction in the case that the width remains in its vacuum value.
Finally, as is shown in comparison of the upper-right with the lower-left,
the width dependence becomes weaker if we choose larger $\xi$. Its
consequence will be discussed in the next section.

\section{Charmonium in hot gluonic matter}
\label{sec:GP}

In this section, we present the result of the analysis for the hot
gluonic matter.
The parameters of the theory are $\alpha_{\text{s}}$ and $m_c$.
Hereafter, they are set to 0.21 and 1.24 GeV at $\xi=1$, that are taken
from \cite{Klingl_PRL82}, respectively.

 \begin{figure}[ht!]
 \includegraphics[width=3.375in]{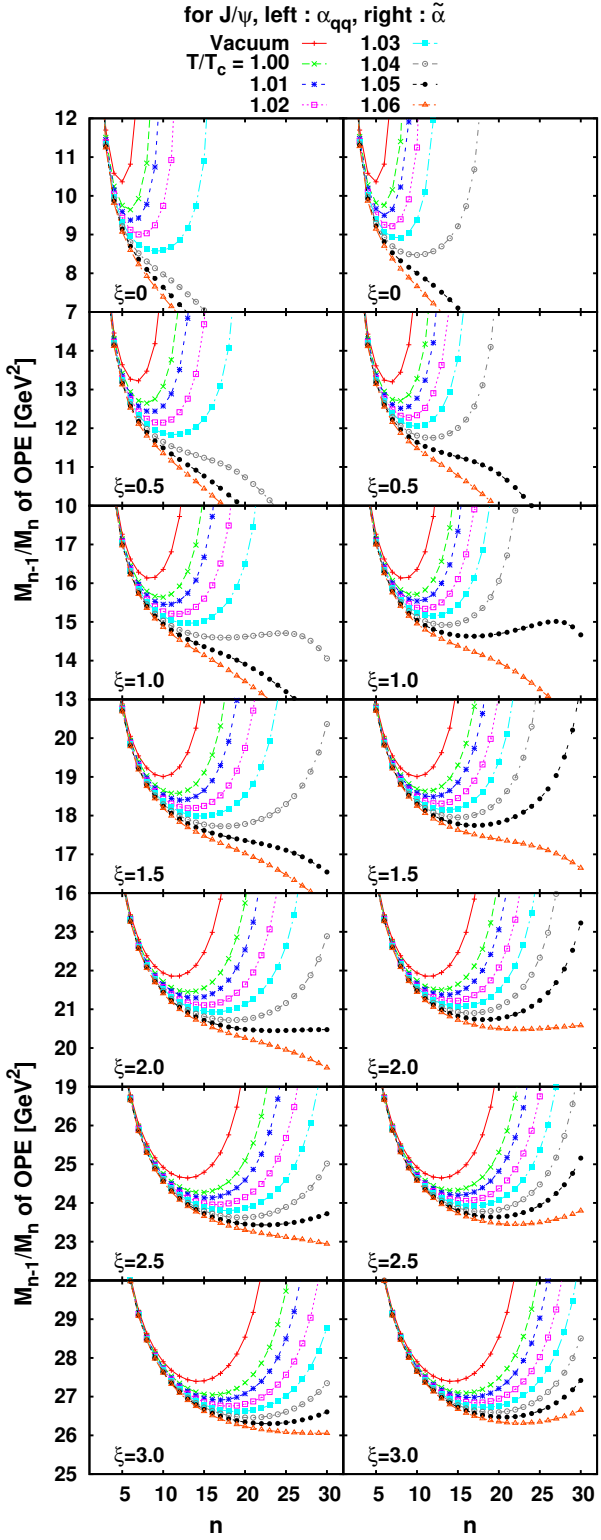}
 \caption{(Color online) Moment ratio for the OPE side for the vector channel
 ($J/\psi$). Each panels show different $\xi$ and coupling constant
 case. The symbols stand for different temperature.}
 \label{fig:ope_jpsi}
 \end{figure}

 \begin{figure}[ht!]
 \includegraphics[width=3.375in]{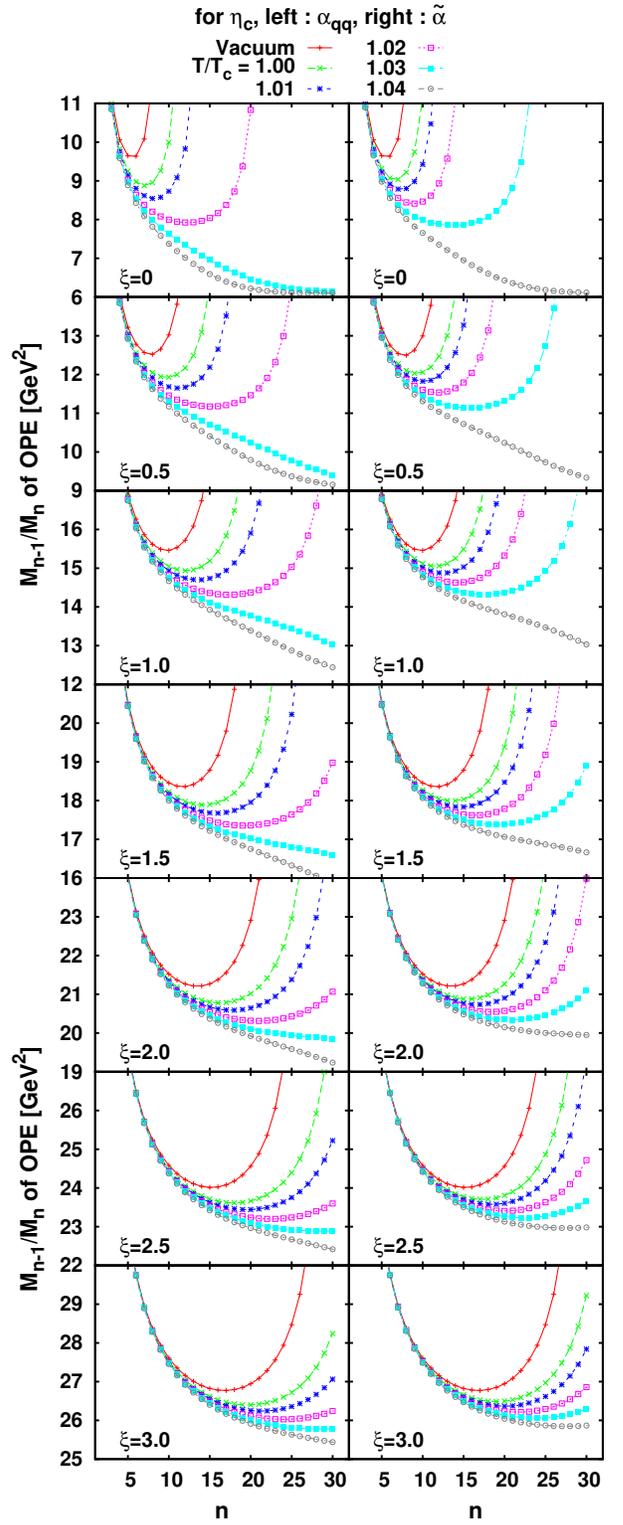}
 \caption{(Color online) Same as Fig.~\ref{fig:ope_jpsi}, but for the
  pseudoscalar channel ($\eta_c$).}
 \label{fig:ope_etac}
 \end{figure}

We begin with fixing $n$ such that the moment ratio of the OPE side
takes its minimum value for each temperature. As briefly mentioned
before, we need to choose moderately large $n$ so that contribution from
excited states and continuum can be sufficiently suppressed. Therefore,
this ratio should approach a constant value at adequately large
$n$. However, in the OPE side contribution from higher dimensional
operators will be important at large $n$.
As such $n$ value that the moment ratio takes its minimum value, pole
dominance and truncation of the OPE are valid and the ratio is close to
the real asymptotic value, as have been extensively studied in the
vacuum case \cite{Reinders_NPB186}.

We display the moment ratio for the OPE side
[Eq.~\eqref{eq:moment_ope_medium}] in Figs.~\ref{fig:ope_jpsi} and
\ref{fig:ope_etac}
with the gluon condensates shown in Fig.~\ref{fig:gc}.

\begin{figure}[ht!]
 \includegraphics[width=3.375in]{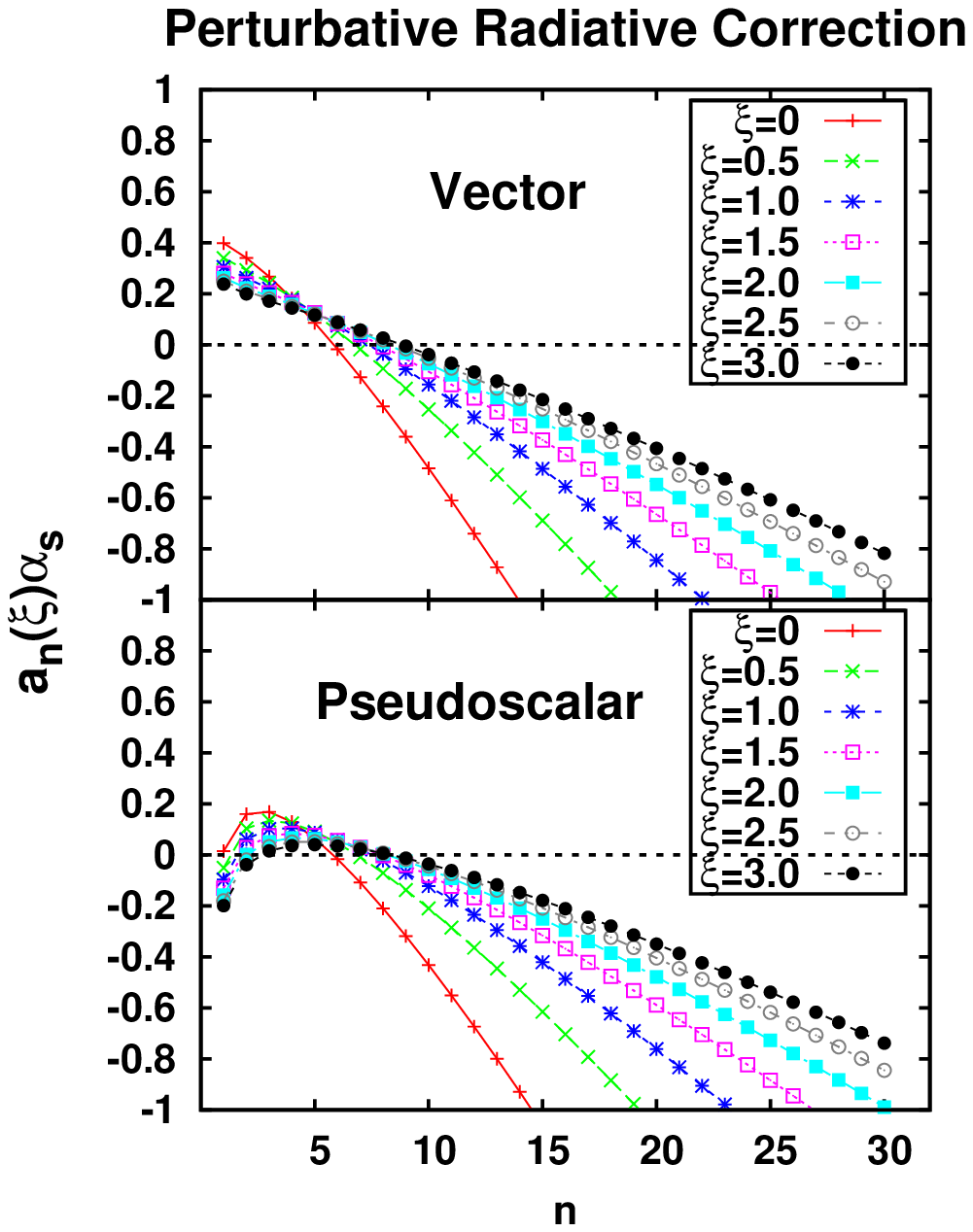}
 \caption{(Color online) Radiative correction term
 $a_n(\xi)\alpha_{\text{s}}$ in the OPE. Upper panel shows the vector
 case and lower one shows the pseudoscalar case.}
 \label{fig:ope_cof_a}
\end{figure}

\begin{figure}[ht!]
 \includegraphics[width=3.375in]{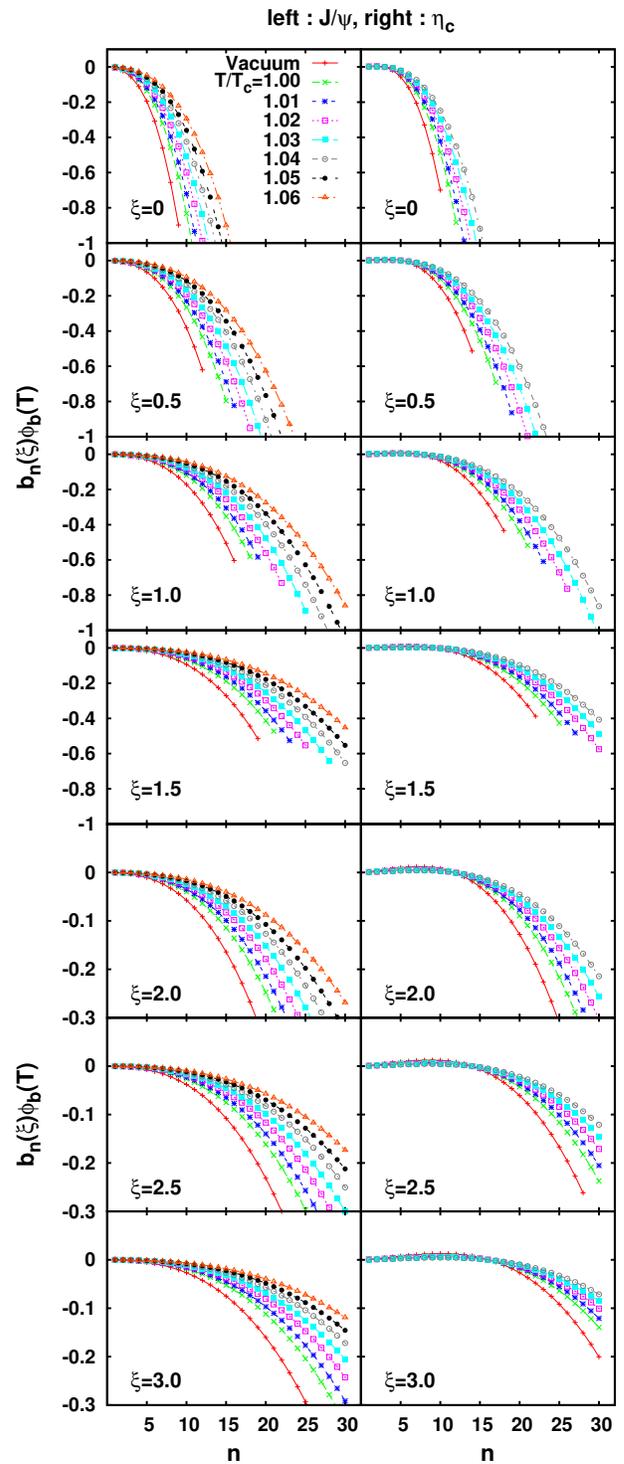}
 \caption{(Color online) Scalar condensate term
 $b_n(\xi)\phi_{\text{b}}$. Left and
 right column stand for the vector and the pseudoscalar case,
 respectively. Symbols denote different temperature cases.}
 \label{fig:ope_b}
\end{figure}

\begin{figure}[ht!]
 \includegraphics[width=3.375in]{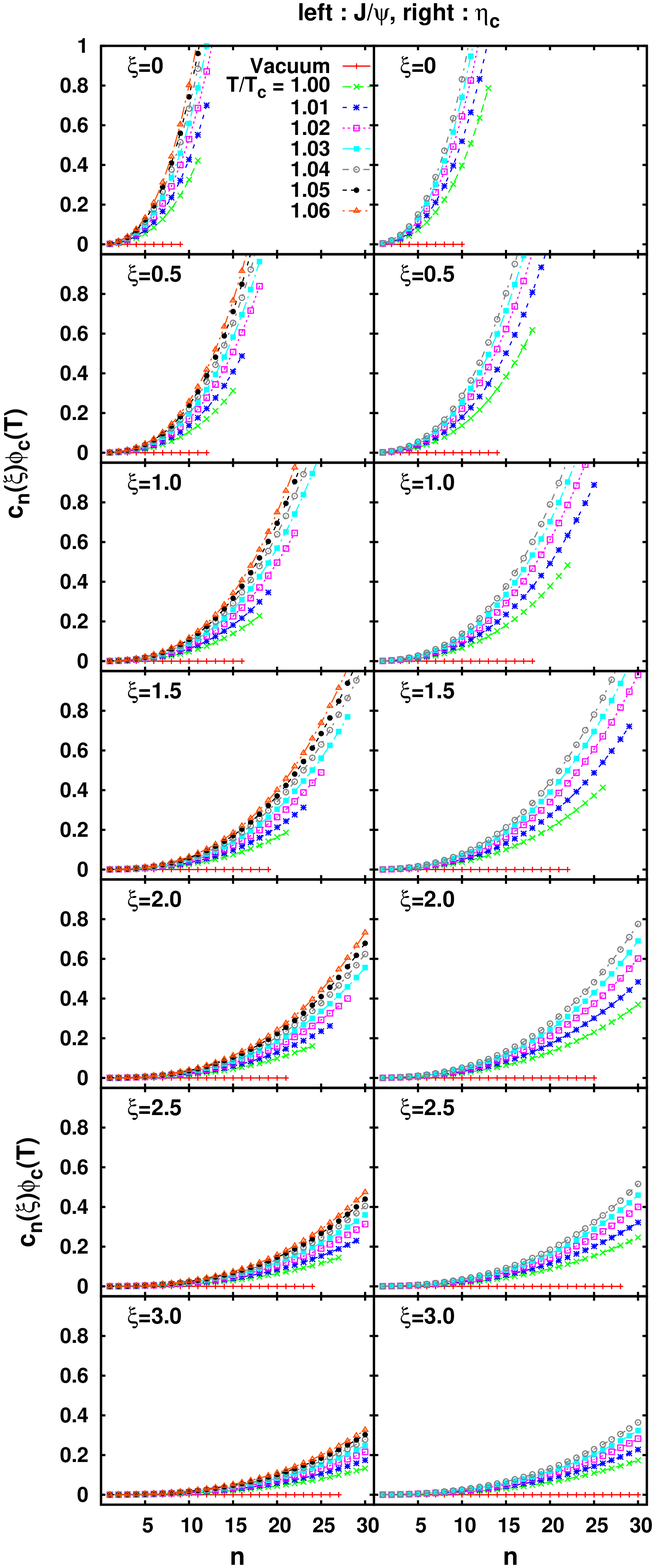}
 \caption{(Color online) Twist-2 condensate term $c_n(\xi)\phi_{\text{c}}$ with
 $\alpha_{\text{qq}}$.}
 \label{fig:ope_c}
\end{figure}

Figure \ref{fig:ope_jpsi} shows the moment
ratio for the vector channel. The left and right column show the case in
which we use $\alpha_{\text{qq}}$ and $\tilde{\alpha}$, respectively.
Comparing different $\xi$ cases, we can see that the stability of the moment
becomes better as $\xi$ increases. But the values of $n$ which give the
stability to the moment ratio also becomes larger. As previously
reported in \cite{Morita_jpsiprl}, the stability is only achieved
near $T_{\text{c}}$ and the stronger coupling, which is
$\alpha_{\text{qq}}$ in this
temperature region, gives worse stability. By increasing $\xi$, we can
improve the stability a little. While it is achieved only up to
1.04$T_{\text{c}}$ for $\xi=0$, the moment ratio remains stable up to
1.06$T_{\text{c}}$ for $\xi=3$.

We can see the similar situation in the pseudoscalar channel depicted in
Fig.~\ref{fig:ope_etac}. However, the moment
ratio is less stable than the vector case. In the pseudoscalar case,
even the best case (using $\tilde{\alpha}$ and $\xi=3$) can stabilize
the moment ratio only up to $1.04T_{\text{c}}$.

Note that the lack of stability does not necessarily mean dissociation of
the charmonia.
The reason of such instability can be clearly seen in the each terms of
the OPE [Eq.~\eqref{eq:moment_ope_medium}], of which each term must be
much less than
unity for convergence. These terms are displayed in
Figs.~\ref{fig:ope_cof_a}-\ref{fig:ope_c}. We can see that
all the coefficients grow up with $n$. An important feature in all the
coefficients
is that increasing $\xi$ clearly keeps their value smaller. Among these
three, only $c_n(\xi)\phi_{\text{c}}$ always has positive sign and its
magnitude increases with temperature. These two features are opposite to
$b_n(\xi)\phi_{\text{b}}$, in which the sign is always negative and the
value seems to approach to zero as temperature increases. In comparing the
two channels, one finds that there are no significant differences.
Hence, the stability will be determined by a delicate balance between
coefficients and its breakdown will be caused by rapid increase of
$c_n(\xi)\phi_{\text{c}}$.

\begin{table}[h!]
\caption{List of $n$ values at which the moment ratio takes minimum values.}
\label{tbl:stability}
\begin{ruledtabular}
\begin{tabular}{c|c|ccrccccccc}
 J& $\alpha_{\text{s}}(T)$ & \multicolumn{1}{c}{$\xi$} &
 \multicolumn{1}{c}{Vac.} & \multicolumn{1}{c}{$\frac{T}{T_c}$=1} &
 \multicolumn{1}{c}{1.01} & \multicolumn{1}{c}{1.02} &
 \multicolumn{1}{c}{1.03} & \multicolumn{1}{c}{1.04} &
 \multicolumn{1}{c}{1.05} & \multicolumn{1}{c}{1.06} \\ \hline

\multirow{14}{*}{$J/\psi$} & \multirow{7}{*}{$\alpha_{\text{qq}}$} & 0 & 5 & 6 & 6 & 7 & 9 & N/A & N/A & N/A \\
 &  & 0.5 & 7 & 8 & 9 & 10 & 11 & N/A & N/A & N/A \\
 &  & 1 & 8 & 10 & 10 & 12 & 13 & N/A & N/A & N/A \\
 &  & 1.5 & 10 & 11 & 12 & 13 & 15 & 17 & N/A & N/A \\
 &  & 2 & 11 & 13 & 14 & 15 & 16 & 18 & 23 & N/A \\
 &  & 2.5 & 13 & 15 & 15 & 16 & 18 & 20 & 22 & N/A \\
 &  & 3 & 14 & 16 & 17 & 18 & 19 & 21 & 23 & 29 \\ \cline{2-11}
 & \multirow{7}{*}{$\tilde{\alpha}$} & 0 & 5 & 6 & 6 & 7 & 8 & 10 & N/A
				     & N/A \\
 &  & 0.5 & 7 & 8 & 8 & 9 & 10 & 12 & N/A & N/A \\
 &  & 1 & 8 & 10 & 10 & 11 & 12 & 13 & 17 & N/A \\
 &  & 1.5 & 10 & 11 & 12 & 13 & 14 & 15 & 17 & N/A \\
 &  & 2 & 11 & 13 & 13 & 14 & 15 & 16 & 18 & 23 \\
 &  & 2.5 & 13 & 14 & 15 & 16 & 17 & 18 & 19 & 22 \\
 &  & 3 & 14 & 16 & 16 & 17 & 18 & 19 & 21 & 23 \\ \hline
\multirow{14}{*}{$\eta_c$} & \multirow{7}{*}{$\alpha_{\text{qq}}$} & 0 & 6 & 7 & 8 & 12 & N/A & N/A & N/A & N/A \\
 &  & 0.5 & 8 & 10 & 11 & 15 & N/A & N/A & N/A & N/A \\
 &  & 1 & 10 & 12 & 14 & 17 & N/A & N/A & N/A & N/A \\
 &  & 1.5 & 12 & 14 & 16 & 19 & N/A & N/A & N/A & N/A \\
 &  & 2 & 14 & 16 & 18 & 21 & N/A & N/A & N/A & N/A \\
 &  & 2.5 & 15 & 18 & 20 & 22 & 29 & N/A & N/A & N/A \\
 &  & 3 & 17 & 20 & 21 & 24 & 29 & N/A & N/A & N/A \\ \cline{2-11}
 & \multirow{7}{*}{$\tilde{\alpha}$} & 0 & 6 & 7 & 7 & 9 & 14 & N/A & N/A & N/A \\
 &  & 0.5 & 8 & 9 & 10 & 12 & 16 & N/A & N/A & N/A \\
 &  & 1 & 10 & 12 & 13 & 14 & 18 & N/A & N/A & N/A \\
 &  & 1.5 & 12 & 14 & 15 & 16 & 19 & N/A & N/A & N/A \\
 &  & 2 & 14 & 16 & 17 & 18 & 21 & N/A & N/A & N/A \\
 &  & 2.5 & 15 & 17 & 18 & 20 & 22 & 28 & N/A & N/A \\
 &  & 3 & 17 & 19 & 20 & 22 & 24 & 28 & N/A & N/A \\
\end{tabular}
\end{ruledtabular}
\end{table} %

Now we proceed to the determination of mass and width.
The values of $n$ are listed in
Table~\ref{tbl:stability}. Note that the stability achieved
at the highest temperature is ambiguous in some cases; for example, $J/\psi$ of
$\xi=2$ with $\alpha_{\text{qq}}$ case is stable at
$T/T_{\text{c}}=1.05$ with $n=22$. However, as seen in
Fig.~\ref{fig:ope_jpsi}, the moment ratio is almost constant in such
large $n$ region and never rises up as lower temperature cases do. Such a
vague stability is also seen in other cases. Hence, we note that mass
and width values evaluated on the basis of such a stability are less reliable
in the analyses below.
\begin{figure}[ht!]
 \includegraphics[width=3.375in]{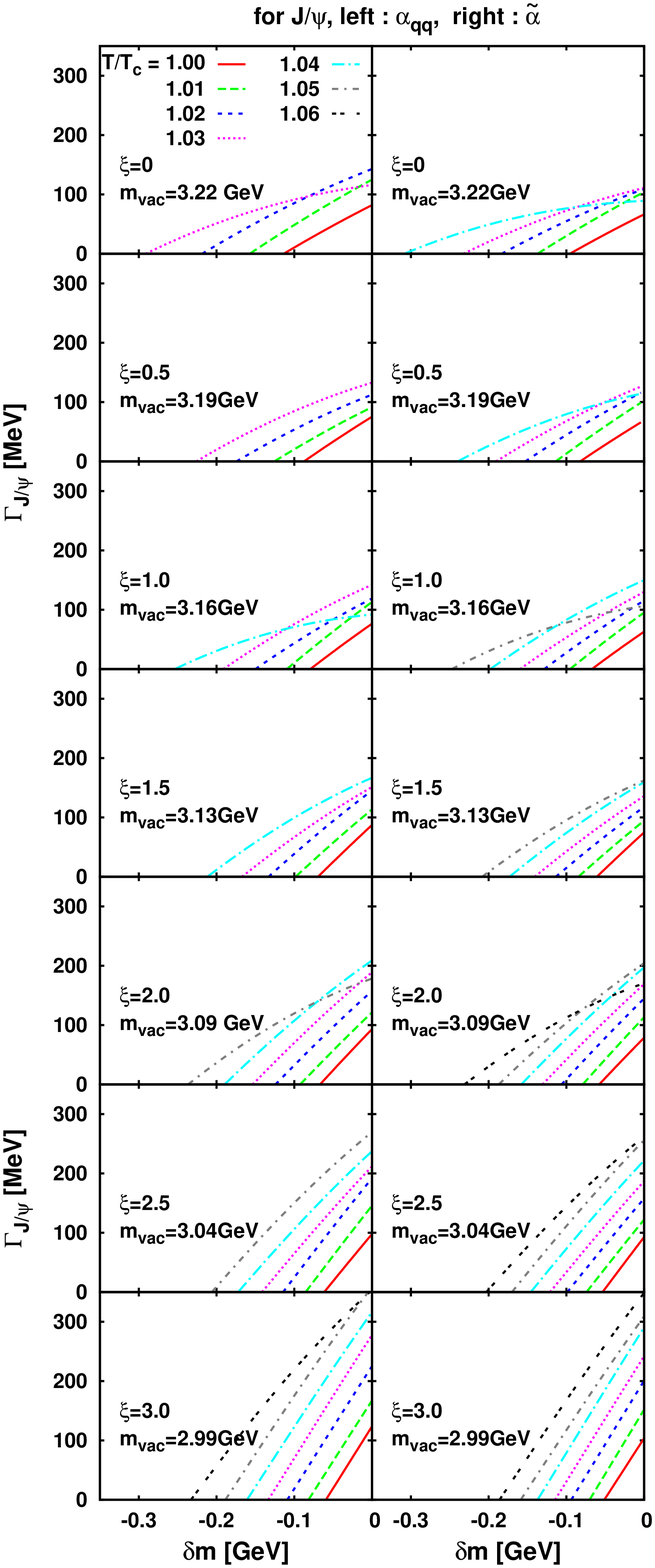}
 \caption{(Color online) Relation between mass shift $\delta m=m_{\text{vacuum}}-m$ and
 width $\Gamma$ for $J/\psi$. As in
 Figs.~\ref{fig:ope_jpsi} and \ref{fig:ope_etac}, each figure shows
 different $\xi$ case for two cases of the coupling constant,
 $\alpha_{\text{qq}}$ and $\tilde{\alpha}$.}
 \label{fig:resultjpsi1}
\end{figure}

\begin{figure}[ht!]
 \includegraphics[width=3.375in]{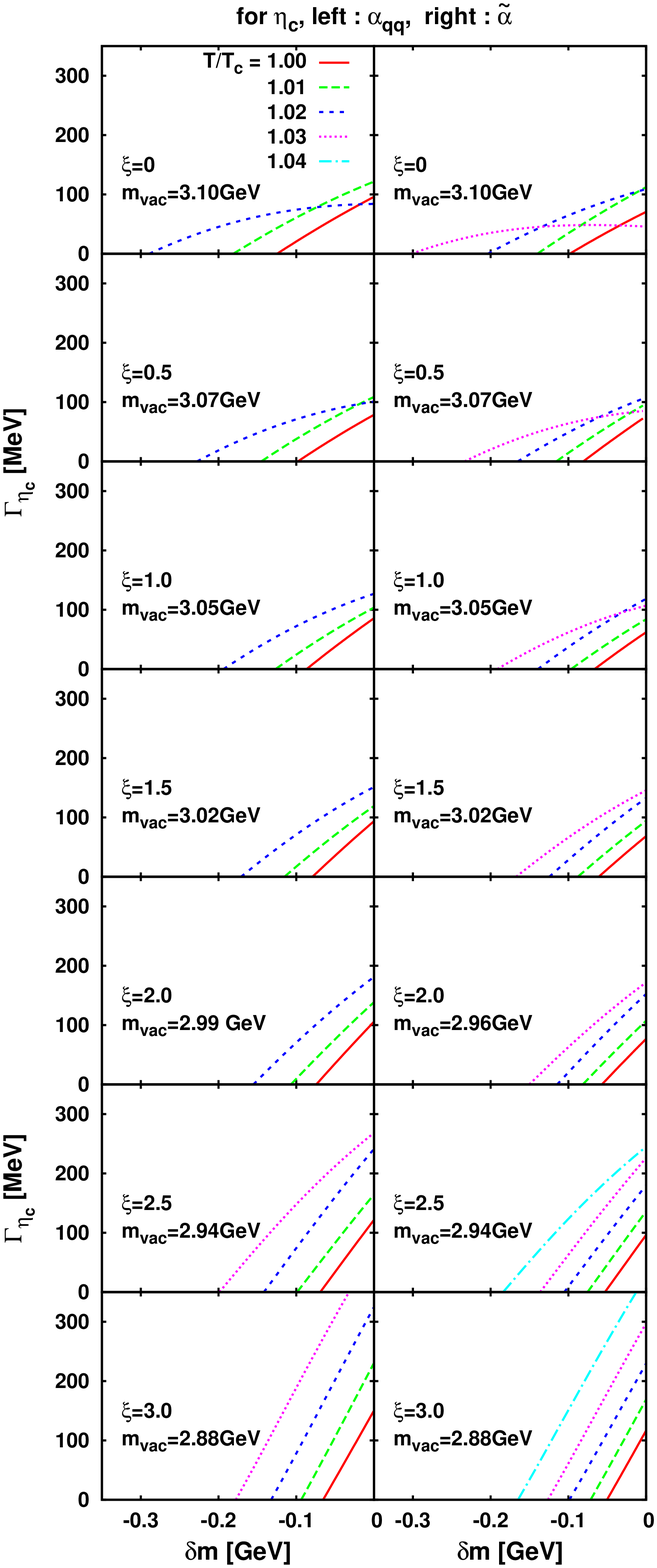}
 \caption{(Color online) Same as Fig.~\ref{fig:resultjpsi1}, but for $\eta_c$.}
 \label{fig:resultetac1}
\end{figure}

\begin{figure}[ht!]
 \includegraphics[width=3.375in]{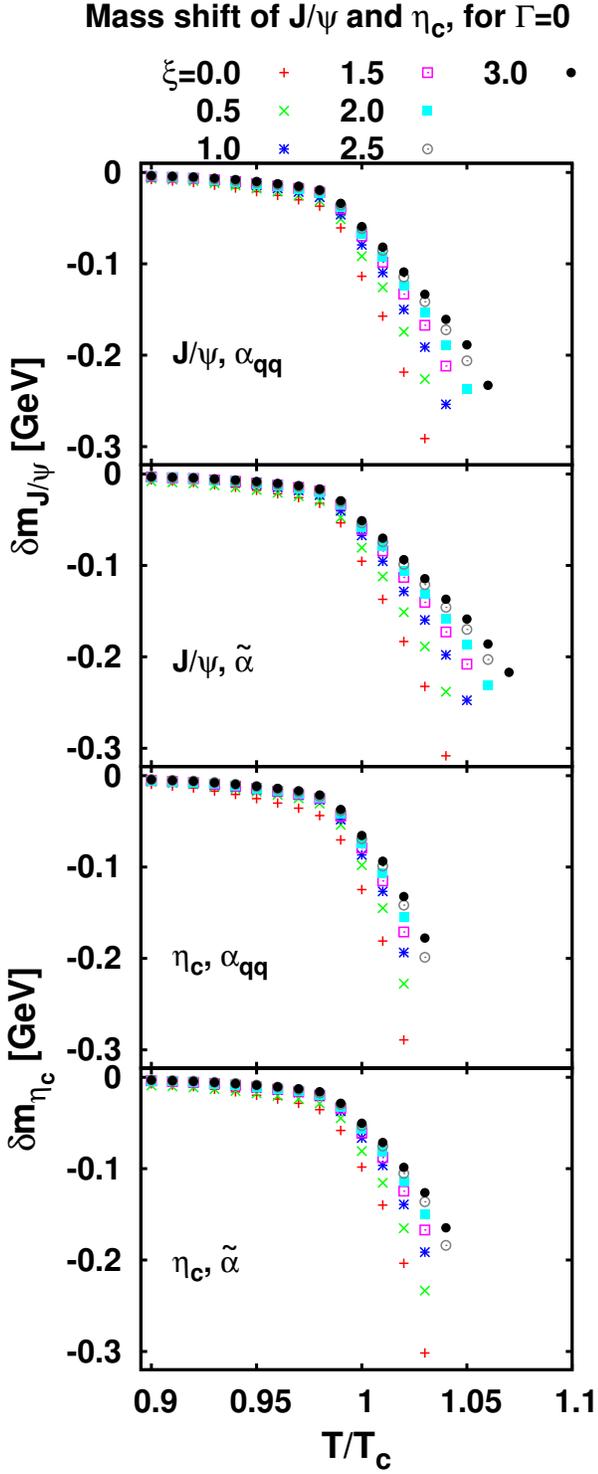}
 \caption{(Color online) Temperature dependence of the masses in the $\Gamma\rightarrow 0$
 limit. Symbols stand for different $\xi$ values.}
 \label{fig:T-m}
\end{figure}

\begin{figure}[ht!]
 \includegraphics[width=3.375in]{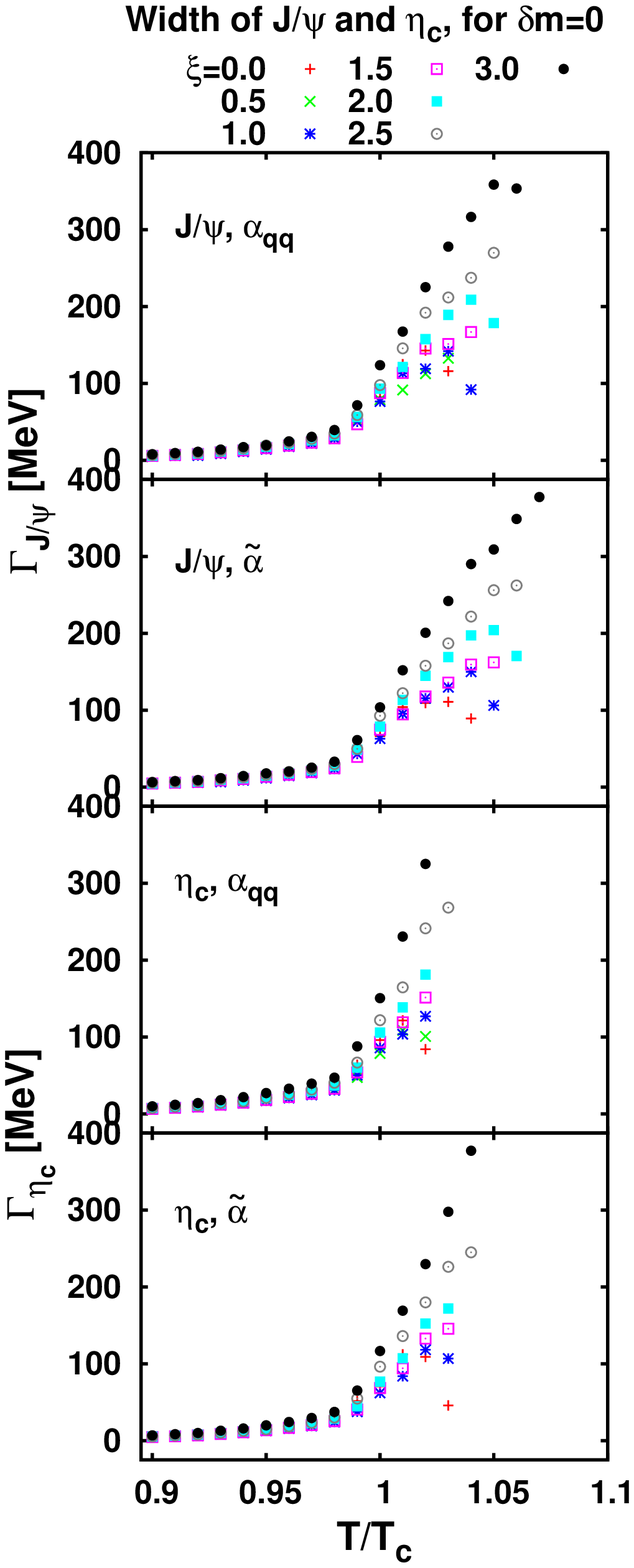}
 \caption{(Color online) Temperature dependence of the widths in the $m \rightarrow 0$
 limit. Symbols are the same as in Fig.~\ref{fig:T-m}.}
 \label{fig:T-g}
\end{figure}

Once $n$ and $\xi$ are fixed, we can compute the mass and the width by
making use of Eq.~\eqref{eq:sumrule}. For a fixed moment ratio of the
OPE side, we firstly compute the mass in the limit of
$\Gamma\rightarrow 0$ using Eq.~\eqref{eq:zerowidthlimit}. By virtue of
the monotonic behavior of the moment ratio of the phenomenological side
shown in Fig.~\ref{fig:phen}, we can safely calculate the mass in the
case of finite width by numerically solving Eq.~\eqref{eq:sumrule} with
Eq.~\eqref{eq:BW}.

We plot the relation between the mass shift and the width at various
temperatures in Figs.~\ref{fig:resultjpsi1} and \ref{fig:resultetac1}.
We can see the almost linear behavior of the width as a function of the
mass shift. Note that the vacuum mass differs for different $\xi$. We do
not perform fine tuning of the parameters so that the real vacuum mass is
reproduced. Although there are some exceptions for the linear relation,
especially small $\xi$ and high temperature cases, these come from the
vague stability we mentioned before. Hence, we can conclude that
the mass shift and the width have the linear relationship as far as QCD
sum rules properly work. The other
important aspect is temperature dependence of the mass shift and the
width. We cannot know how the mass and the width behave in the real
situation, since we cannot simultaneously determine both of the mass and
the width within the current framework only. Here, we investigate two
extreme cases; $\Gamma\rightarrow 0$
limit and $\delta m \rightarrow 0$ limit.

The results are shown in Figs.~\ref{fig:T-m} and \ref{fig:T-g}.
In these figures, we plot the results of $T > 0.9T_{\text{c}}$.
Figure \ref{fig:T-m} shows the remarkable behavior of the mass shift;
The mass does not change up to $T \sim T_{\text{c}}$ but it suddenly
begins to decrease across $T_{\text{c}}$. This fact clearly reflects
the temperature dependence of the gluon condensates which represent the
phase transition. Above $T_{\text{c}}$, the mass decreases with
temperature almost linearly.
This feature is common for both $J/\psi$ and $\eta_c$.
Though small $\xi$ results, especially $\xi=0$, show more rapid
decrease, the curves become almost parallel among large $\xi$ results,
as a consequence of the better stability. From the nature of the
phenomenological side shown in Fig.~\ref{fig:phen}, this case
corresponds to the maximum mass shift. The mass shift shows $\sim$ 50
MeV reduction from vacuum to $T_{\text{c}}$, and it increases
additionally by $\sim$ 20-50 MeV as
temperature rises by 0.01$T_c$. Consequently , it becomes 100-300 MeV at
$T=1.04T_{\text{c}}$.

Similarly, Fig.~\ref{fig:T-g} shows that the width begins to increase with
temperature across $T_{\text{c}}$ if no mass shift takes place. This also shows
almost linear dependence on temperature above $T_{\text{c}}$.
Though some
exceptions can be seen in the small $\xi$ results, which are also
indicated in Figs.~\ref{fig:resultjpsi1} and \ref{fig:resultetac1}, these
behaviors come from the vague stability appearing as too large $n$ in
Table~\ref{tbl:stability}. Hence, we can conclude that the width
increases linearly with temperature above $T_{\text{c}}$ if the mass
remains unchanged. Since we
did not do fine tuning of the parameters for each $\xi$, the values of
mass shift and width differ for different $\xi$. However, the qualitative
features do not depend on $\xi$ where the stability is reliable. This
shows robustness of our analysis. A realistic change at each temperature
should be a combined decrease in mass and increase in width, whose
values are smaller than their maximal changes obtained here. However, to
determine the realistic combination, we need to have an additional
constraint between the changes in the width and the mass, or input the
thermal width from another calculation.\footnote{See
Ref.~\cite{lee_morita_stark} for a recent investigation.}
\begin{figure}[ht]
 \includegraphics[width=3.375in]{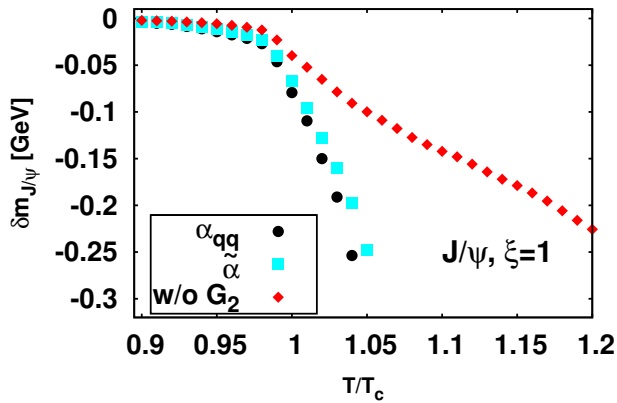}
 \caption{(Color online) $\delta m$ of $J/\psi$ without twist-2 term
 $G_2$. See text for detail.}
 \label{fig:withoutg2}
\end{figure}
From Fig.~\ref{fig:gc}, one may think that the dominant contribution to the
change of mass and/or width is the scalar gluon condensate which exhibits
sudden decrease around $T_{\text{c}}$. However, $G_2$ has also similar
behavior since it relates to the entropy density. Though the value of
$G_2$ around $T_{\text{c}}$ is smaller than $G_0$ because of prefactor
$\alpha_s/\pi$, the relative contribution to the moment becomes larger as $T$
increases. In order to see the contribution clearly, We show the mass
shift of $J/\psi$ without $G_2$ term for $\xi=1$ together with the two
different coupling cases in Fig.~\ref{fig:withoutg2}. We can see that almost half of the mass shift
is caused by decrease of $G_2$. Clearly larger $G_2$ value in which
$\alpha_{\text{qq}}$ is adopted as coupling constant leads to larger
mass shift. At $T=1.04T_{\text{c}}$, $\alpha_{\text{qq}}$ is about 0.1
larger than $\tilde{\alpha}$. This difference makes the mass shift 30
MeV larger in the $\xi=1$ case. Unfortunately present analysis is
limited to the temperature region around $T_{\text{c}}$, the role of
twist-2 term will become more important at higher temperature.

\section{Nuclear matter}
\label{sec:NM}

In this section, we analyze change of mass and width of the charmonium
induced by nuclear medium with the same framework that was implemented in the
previous section. Here, we use Eqs.~\eqref{eq:gc_nm} and \eqref{gc2_nm}
instead of Eqs.~\eqref{eq:phib_qgp} and \eqref{eq:phic_qgp},
respectively. With the common parameter set, the condensates are
$\phi_{\text{b}}=1.74\times 10^{-3}$ for vacuum, $1.64\times 10^{-3}$
for the nuclear matter, and $\phi_{\text{c}}=-1.28\times 10^{-5}$.

As previously shown in Ref.~\cite{Klingl_PRL82}, the change of mass,
which is identical to the change of the moment ratio of the OPE side
[Eq.\eqref{eq:zerowidthlimit}], is
not as large as in the hot gluonic matter case. Thus we do not have to
worry about the
stability of the OPE. Nevertheless, increasing $\xi$ improves the
validity of the OPE. We will show the results for $0 \leq \xi \leq 3$ as
well as in the hot gluonic matter case to show the robustness and the
consistency of the calculation.

\begin{table}[ht!]
 \caption{List of $n$ which stabilize the moment ratio for the nuclear matter}
 \label{tbl:nm}
 \begin{ruledtabular}
  \begin{tabular}{cccccccc}
   channel&$\xi=0$&$\xi=0.5$&$\xi=1$&$\xi=1.5$&$\xi=2$&$\xi=2.5$&$\xi=3$
   \\ \hline
   $J/\psi$&5&7&9&10&12&13&14  \\
   $\eta_c$&6&8&10&12&14&15&17 \\
  \end{tabular}
 \end{ruledtabular}
\end{table}

\begin{figure}[ht!]
 \includegraphics[width=3.375in]{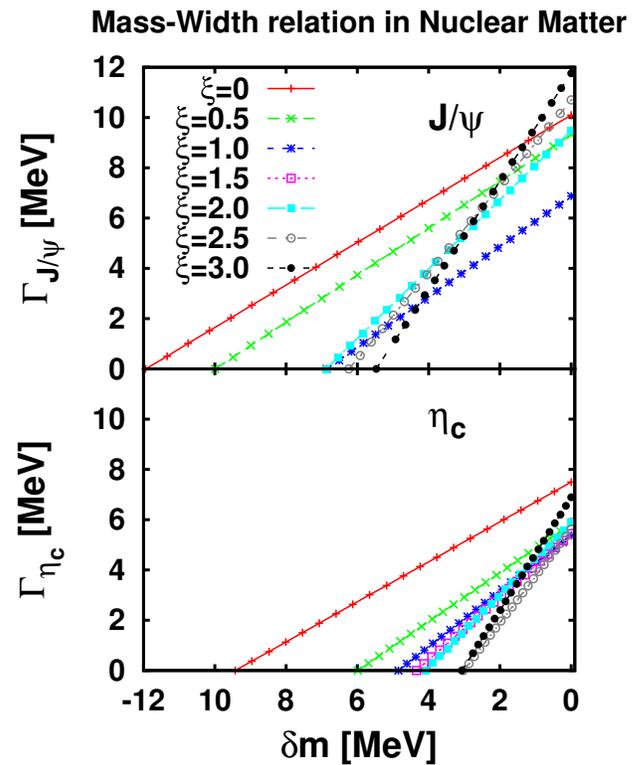}
 \caption{(Color online) Relation between mass shift and width in the
 nuclear matter.}
 \label{fig:dm-g_nm}
\end{figure}

We list values of $n$ in which the moment ratio of the OPE side for
nuclear medium becomes minimum in Table~\ref{tbl:nm}. The vacuum case
has been already shown in Table~\ref{tbl:stability}. Comparing these
two cases, we can see that the values of $n$ are the same except for
a few exception in the $J/\psi$ case, by virtue of the small shift of
the gluon condensates in the nuclear matter. In such exceptional cases,
difference of the values of the moment ratio from the same $n$ value
case with the vacuum is almost negligible. \textit{i.e.}, the moment
ratio is almost constant around these $n$.

We plot the width $\Gamma$ as a function of the mass shift $\delta m$ in
Fig.~\ref{fig:dm-g_nm} as well as in the GP case. In both $J/\psi$ and
$\eta_c$ cases, smaller $\xi$ than 1.5 show larger mass shift and width
broadening but larger $\xi$ results agree each other. From the stability
argument, larger $\xi$ results will be more reliable. Then, possible
mass shifts are maximally -7 MeV for $J/\psi$ and -4 MeV $\eta_c$ while
maximum widths are 10 MeV for $J/\psi$ and 6 MeV for $\eta_c$.

\section{Discussion and Summary}
\label{sec:summary}

\begin{table*}[!ht]
 \caption{Parameters and results in charmonium productions at GSI-FAIR.
 Cross sections and event per day correspond to the case of maximum
 medium width, $\Gamma_{\text{med}}=20$ MeV.}
 \label{tbl:sigma_bw}
 \begin{ruledtabular}
  \begin{tabular}{ccccccc}
   Resonance&$m$[MeV]&$\delta m$[MeV]&$\Gamma_{\text{tot}}$&Final
   State&$\overline{\sigma_{\text{BW}}}$ at peak&Events per day
   \\\hline
   $J/\psi$&3097&-7&93.4keV&$e^+ + e^-$&0.435 pb&7.5 \\
   $\eta_c$&2980&-4&25.5MeV&$e^+ + e^-$&10.7 pb&184 \\
   $\chi_{c0}$&3415&-60&10.4MeV&$J/\psi + \gamma$&18.0 pb&311 \\
   $\chi_{c1}$&3511&-60&0.89MeV&$J/\psi + \gamma$&4.5 pb&78 \\
   $\chi_{c2}$&3556&-60&2.05MeV&$J/\psi + \gamma$&19.8 pb&343 \\
  \end{tabular}
 \end{ruledtabular}
\end{table*}

In Sec.~\ref{sec:GP}, we have shown that
mass decreases suddenly across $T_{\text{c}}$ and the shift reaches
maximally a few hundred MeV above $T_{\text{c}}$ in the hot gluonic matter.
Alternatively, width can also maximally broaden to $\sim$200 MeV.
Although our analysis
cannot determine both of mass and width simultaneously, this is a
notable result which should be examined in the present and future
experiments. In fact, a next to leading order QCD calculation shows
that the thermal width of $J/\psi$ slight above $T_{\text{c}}$ is smaller
than a few 10 MeV \cite{Park,Lee_Thermalwidth}. Hence
a large mass shift will take place. Note that
such a large mass shift has been expected from different points view;
an AdS/QCD analysis shows a sudden drop of mass at the phase transition
\cite{Kim_AdSQCD}. In Ref.~\cite{Kim_AdSQCD}, although the mass begins
to slowly increase at higher temperature, the temperature region
investigated in the present paper corresponds to the critical region.
Sudden reduction of the asymptotic
value of the potential seen in lattice QCD \cite{Karsch04} leads
to lowering of the bound state energy \cite{Wong05}. Recent lattice QCD
calculation based on the maximum entropy method also shows survival of
the peak in the spectral function above $T_{\text{c}}$ \cite{Aarts07} but
the resolution is still insufficient to discuss shift and broadening of
the peak.
Since our results access only near $T_{\text{c}}$, we are still far from
the complete
understanding of the behavior of the charmonium in the deconfined
medium. In the most
plausible picture from the current understanding, charmonia are melting
at very high temperature expected in the early stage of the heavy ion
collisions at RHIC and LHC. Then the pairs of heavy quark and antiquark
form the bound states at a certain temperature which depends on quantum
number. The temperature is expected as $\sim 2T_c$ for $J/\psi$ at RHIC
\cite{Gunji}. After charmonia are produced, they will dissociate by
collisions with partons. If this phase lasts long enough compared to the
inverse of the width, the charmonia can decay in the medium. In fact,
the lifetime of the partonic medium is about 4-5 fm/$c$ in a
hydrodynamic calculation for the central Au+Au collisions at the maximum
RHIC energy \cite{Morita_hydro07}. This will be much longer at LHC. From
Fig.~\ref{fig:resultjpsi1}, we expect $\sim 200$ MeV $J/\psi$ mass
reduction in the case of the small decay width.
This shift is larger than experimental mass resolutions ($\sim$
35 MeV for dielectron channel of PHENIX at RHIC
\cite{PHENIX_Jpsilatest}, 33 MeV for dielectron channel and 75 MeV for
dimuon channel of ALICE at LHC \cite{ALICE_reportII}).

Alternatively, statistical hadronization near phase boundary has been
also examined \cite{Andronic07}. In this case, the number of produced
charmonium will be enhanced if the notable mass shift occurs. For example,
there may be a factor of 2 enhancement for $T=170$ MeV and
$\delta m = -100$ MeV since the enhancement factor is  given by
$e^{-\delta m/T}$, This enhancement might be observed by
comparing particle ratio.

As for the nuclear medium result, we have extended the analysis carried
out in Ref.~\cite{Klingl_PRL82} to the one which takes account of
finite width. We have also shown the results for different $\xi$
values. Since we have given the relation between the mass shift and
width, we can estimate the mass shift in the presence of finite width
effect by considering the dissociation
cross section of the charmonium by nucleon. Provided the Fermi momentum
is $p_{\text{F}}\simeq 250$ MeV and the cross section is
$\sigma_{J/\psi-N}\simeq 2$mb, the decay width
$\Gamma=\langle \sigma_{J/\psi-N} v_{\text{rel}} \rho_N \rangle$ becomes
$\sim 1.3$ MeV for charmonium at rest. The cross section may be smaller,
because the incident momentum is considered to be small and the process
will be near threshold. From this estimate, if we take into account the
broadening of the width, the mass shift becomes slightly smaller, by about
0.5 MeV,
according to the results shown in Fig.~\ref{fig:dm-g_nm}. Therefore, this
justifies the argument in Ref.~\cite{Klingl_PRL82} that the influence of
the decay widths is expected to be small.

\begin{figure}[!ht]
 \includegraphics[width=3.375in]{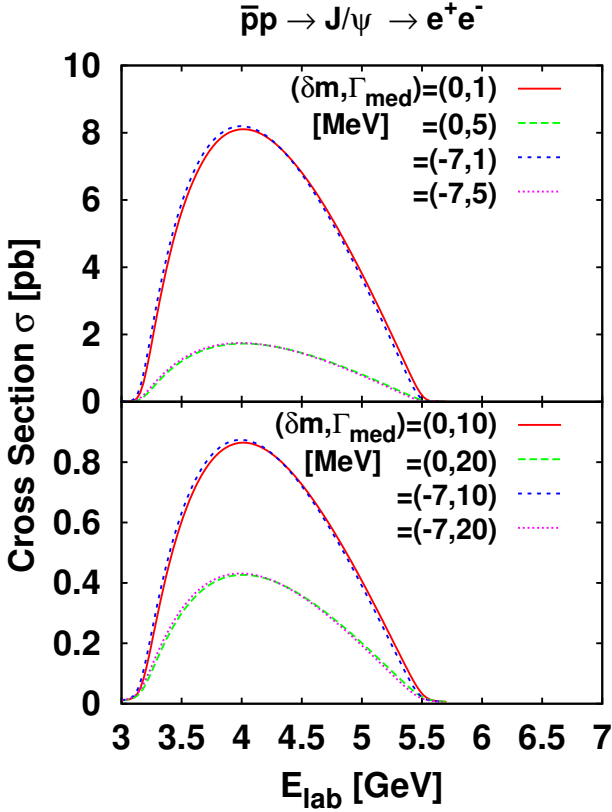}
 \caption{(Color online) Cross section of $J/\psi$ production in $\bar{p}-A$
 collisions. Upper panel shows smaller medium width (1 and 5 MeV) cases
 and lower one shows larger (10 and 20 MeV) cases for with and without
 mass shift.}
 \label{fig:sigma_jpsi}
\end{figure}

\begin{figure}[!ht]
 \includegraphics[width=3.375in]{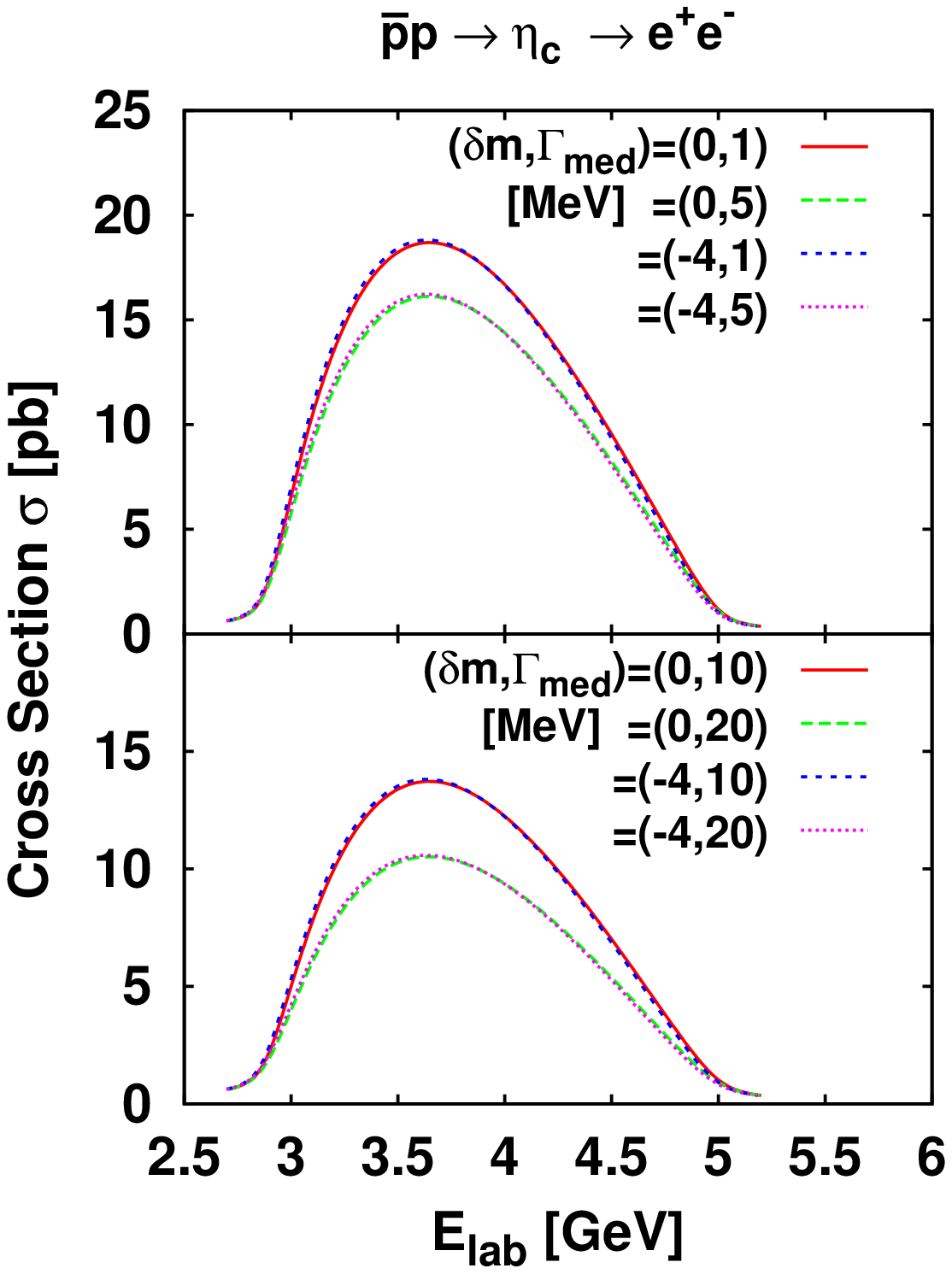}
 \caption{(Color online) Same as Fig.~\ref{fig:sigma_jpsi}, but for $\eta_c$. }
\end{figure}

\begin{figure}[!ht]
 \includegraphics[width=3.375in]{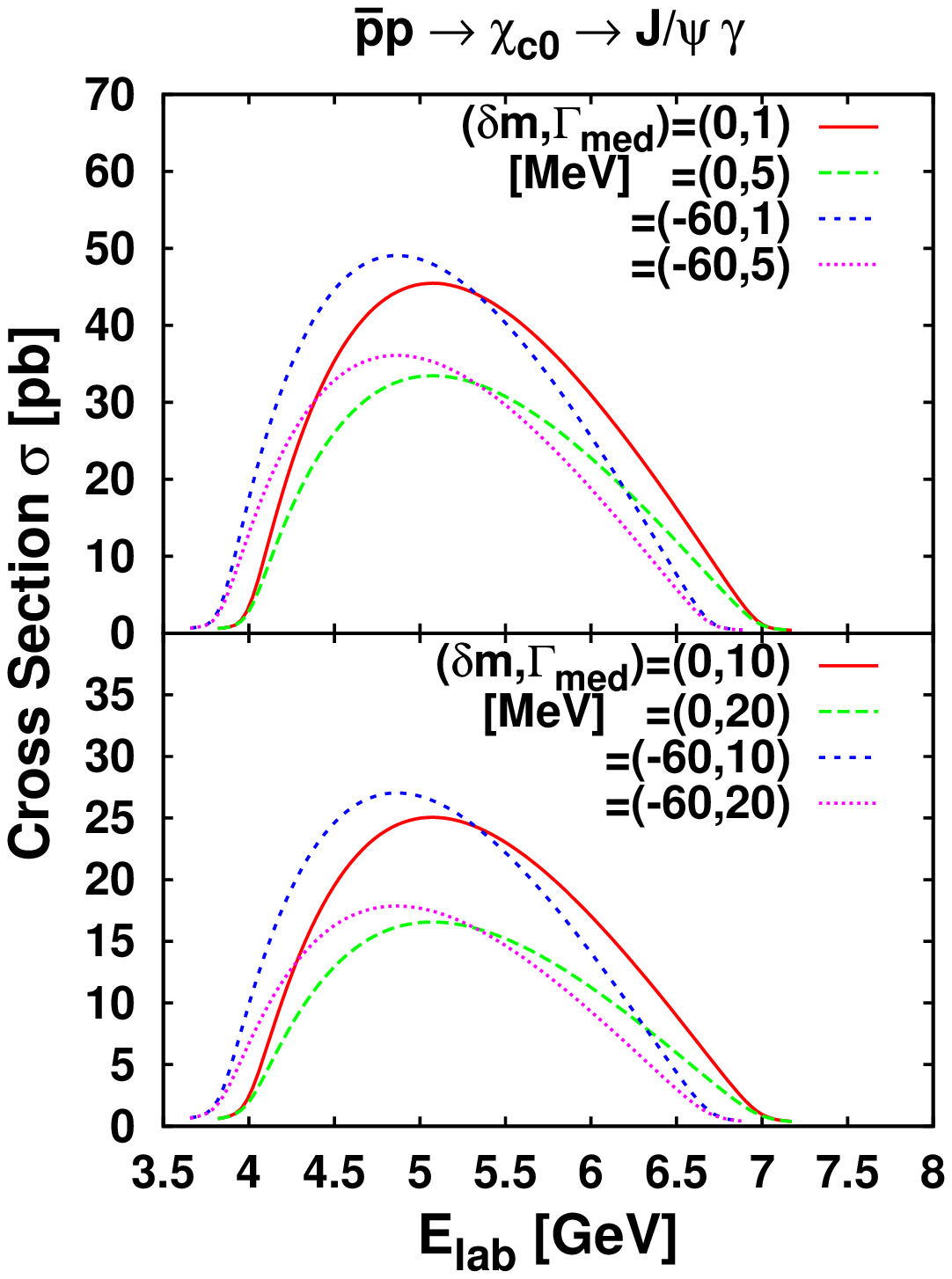}
 \caption{(Color online) Same as Fig.~\ref{fig:sigma_jpsi}, but for $\chi_{c0}$. }
\end{figure}

\begin{figure}[!ht]
 \includegraphics[width=3.375in]{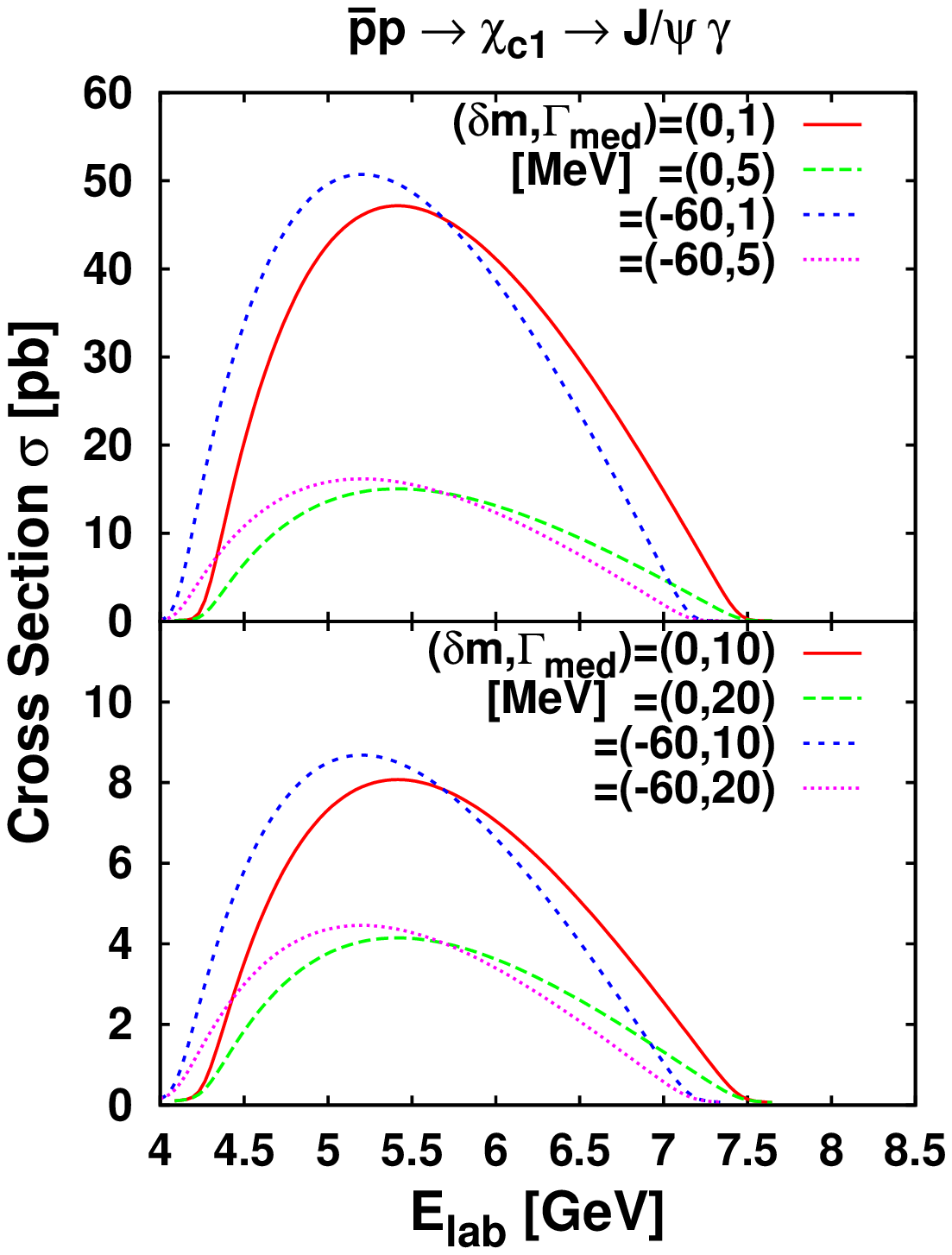}
 \caption{(Color online) Same as Fig.~\ref{fig:sigma_jpsi}, but for $\chi_{c1}$. }
\end{figure}

\begin{figure}[!ht]
 \includegraphics[width=3.375in]{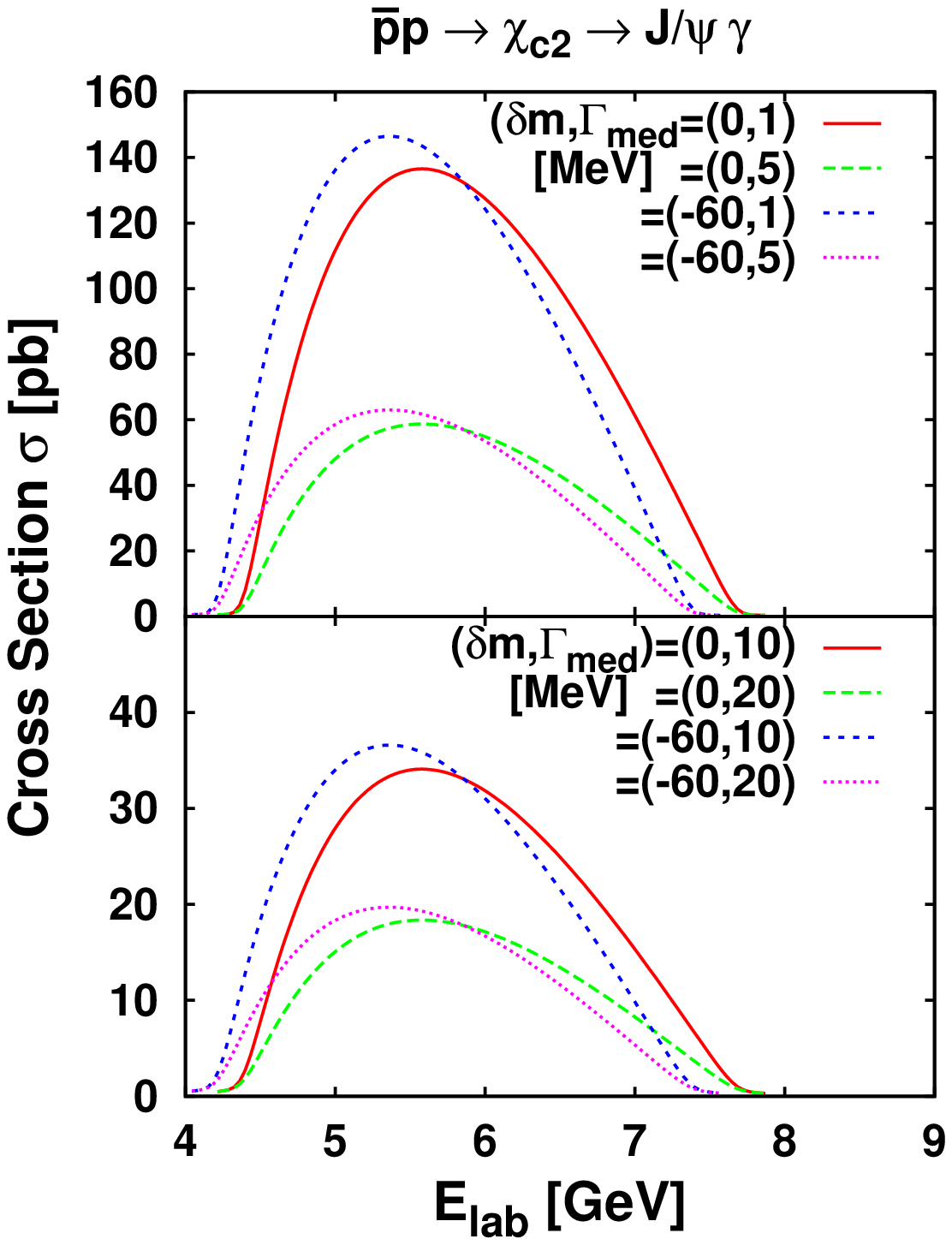}
 \caption{(Color online) Same as Fig.~\ref{fig:sigma_jpsi}, but for $\chi_{c2}$. }
 \label{fig:sigma_chi2}
\end{figure}

The change of spectral properties in the nuclear matter can be
experimentally investigated by Panda experiment at GSI-FAIR in which
incident anti-proton collide with nuclear target. Here we present some
predictions for cross sections of charmonium production through
$\bar{p}p$ annihilation and subsequent
decay into dileptons or radiative decay in the
experiment. We compute the cross sections with the Breit-Wigner formula
\begin{equation}
 \sigma_{\text{BW}}(s)
  =
  \frac{B_{\text{in}}B_{\text{out}}(2J+1)}{(2s_1+1)(2s_2+1)}
  \frac{4\pi}{k^2_{\text{cm}}}
  \frac{s\Gamma^2_{\text{tot}}}{(s-m^2)^2+s\Gamma^2_{\text{tot+med}}},
\end{equation}
where $s$, $k^2_{\text{cm}}$ and $m$ are the Mandelstam variable,
c.m. momentum and mass of charmonium with spin $J$,
respectively. $\Gamma_{\text{tot}}$ is the total decay width of the
charmonium and
$\Gamma_{\text{tot+med}}=\Gamma_{\text{tot}}+\Gamma_{\text{medium}}$.
$B_{\text{in}}$ and $B_{\text{out}}$ are the branching
fraction of the resonance into the entrance and exit channels. $s_i$ is
the spin of the incident particles, which are anti-protons and protons
in the present calculation. Since the target protons are in nucleus, we
have to take the Fermi motion into account for accurate estimation.
We average the Breit-Wigner cross section with respect to target
momentum as
\begin{equation}
 \overline{\sigma_{\text{BW}}} =
  \frac{4}{\rho_0}\int_{0}^{k_{\text{F}}}k^2 \frac{dk d\Omega}{(2\pi)^3}
  \sigma_{\text{BW}}.
\end{equation}

In addition to $J/\psi$ and $\eta_c$, we also calculate cross sections
for $\chi_c$ which are expected to show larger mass shift
$\delta m \simeq -40 \sim -60$ MeV \cite{Lee04}. Parameters in the
calculations are summarized in
Table~\ref{tbl:sigma_bw}. $\Gamma_{\text{medium}}$ is treated as a free
parameter varied from 1 MeV to 20 MeV.

Results of the cross sections as a function of incident anti-proton
energy are shown in Figs.~\ref{fig:sigma_jpsi}-\ref{fig:sigma_chi2}.
We can clearly see that sharp peaks of the resonances disappear.
This is because of the Fermi motion of the target protons in the
nucleus. For example, incident energy to create $J/\psi$ (3097) is
$E_{\text{lab}}=4.17$ GeV, but the fluctuation of the target momentum
makes it possible to create $J/\psi$ with
$3.17 \leq E_{\text{lab}}\leq 5.51$ GeV, in which the minimum and the
maximum $E_{\text{lab}}$ correspond to the target momentum along the
collision axis $p_{2z}=-k_{\text{F}}$ and $k_{\text{F}}$, respectively.
This effect considerably broadens the cross section.
Consequently, one needs no fine tuning of incident proton energy to
produce charmonium. For $J/\psi$ and $\eta_c$, mass shifts are so small
that the peak positions of incident energy do not change.
However,  mass shift of $\chi_c$, $\sim -60$ MeV, is sufficiently large
to show clear shift of the peak in the cross section.
In these calculations, we treat
$\Gamma_{\text{medium}}$ as a free parameter. It is shown that this
parameter affects only on the magnitude of the cross section, which is
larger for smaller change from the vacuum width. Hence, though we cannot
predict both of mass shift and in-medium width, we can obtain
information on both quantities from the experimentally measured cross
sections. We summarized the cross sections and expected event rate at
GSI-FAIR, of which luminosity is expected to be
$2\times 10^{32}\text{cm}^{-2} \text{s}^{-1}$, in last two column of
Table.~\ref{tbl:sigma_bw}. We can see that the expected event rates are
large enough for the mass shift of $\chi_c$ to be observed.

Finally we address possible improvements of this work.
Since we restricted ourselves to the hot medium which
consists of gluons only in the first part of this paper, we should take
the quarks into account for more realistic estimation.
  
To consider the quark effects, first we consider the quark operators
appearing in the OPE side.
We can neglect the light quark contribution to the OPE,  because the
light quark operators appear in the OPE at order $\alpha_s^2(q^2)$: This
is why the light quark condensate can be neglected in the sum rules for
heavy quark system.  On the other hand, thermal heavy quarks that
directly couple to the heavy quark current contribute to the OPE at
leading order. This is different from the heavy quark condensates that
are perturbatively generated from the gluon condensates, and contribute
to the OPE through gluon condensates, whose Wilson coefficients are
calculated in the momentum representation.  The direct thermal quark
contributions are called the scattering terms.  However, similar terms
also appear in the phenomenological side, which also has free charm
quark mode that is not coupled with a light quark in the form of a $D$
meson above $T_{\text{c}}$ as been recently studied in
Ref.~\cite{Dominguez}.  Therefore, the scattering term will cancel out
between the OPE side and the phenomenological side in the deconfined medium.

\begin{figure}[ht]
 \includegraphics[width=3.375in]{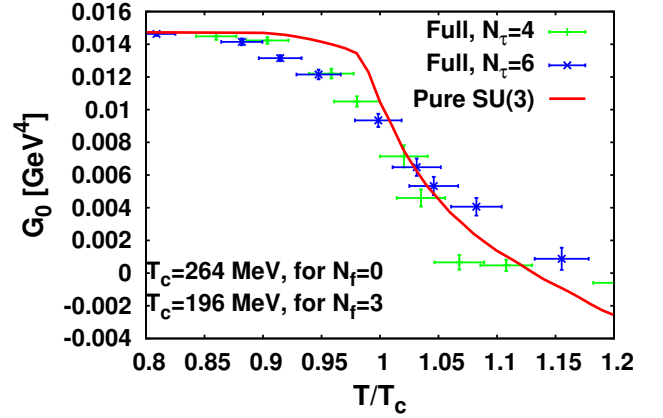}
 \caption{(Color online) Comparison of the scalar
 gluon condensate in the pure gauge theory with the one of full QCD.
 Horizontal errorbars in the full QCD case are drawn by assuming the 2\%
 uncertainty in the conversion from the lattice units to physical temperature
 \cite{cheng08}.}
 \label{fig:e3p_full}
\end{figure}

Second, the gluon condensates themselves can have a different temperature
dependence in the presence of dynamical quarks.  As discussed before,
the important input for the mass and width change is the temperature
dependence of gluon condensates in Fig.~\ref{fig:gc}; in particular the
dominant contribution comes from the temperature dependence of $G_0$.
For that purpose, we note that the trace of the energy momentum tensor
to leading order is given as,

\begin{equation}
 T^{\mu}_\mu = - \left( \frac{11-2/3 N_{\text{f}} }{8} \right)
  \left\langle \frac{\alpha_{\text s}}{\pi} G^{a}_{\mu\nu}G^{a\mu\nu}
  \right\rangle + \sum_q m_q \langle \bar{q}q \rangle. \label{eq:total_trace}
\end{equation}
Therefore, we start from the lattice calculation of the trace of the
energy momentum tensor for the full QCD with realistic quark masses
given in Ref.~\cite{cheng08}.  Then, we subtract the
fermionic part of the trace anomaly, which was also shown in the
literature, from the total. Next, we divide the result for the relevant
prefactor with $N_{\text{f}}=3$ multiplying the gluon condensate as given in
Eq.~\eqref{eq:total_trace}.   Since the critical
temperature $T_c$ differs, we compared it as a function of $T/T_c$ in
which $T_c=196$ MeV for the full QCD case~\cite{cheng08}.
As can be seen in  Fig.~\ref{fig:e3p_full}, the magnitude of the
resulting change near the critical temperature are remarkably similar
between the full and pure gluon QCD; although the slope at $T_c$ is
milder for full QCD as a consequence of rapid cross over transition
instead of the first order phase transition.  
 Since the change of the condensate sets in at a lower $T/T_c$ in the full
QCD case, the mass and width of charmonia might start varying at a lower
temperature in the realistic case than in the pure glue theory. As for
the twist-2 condensates, results
will not be affected so much by taking into account the fermionic part
since the effect will be small at this temperature region. Therefore we believe
our main argument and the quantitative result will not be alter even in the realistic situation.

It is also important
to study change of $\chi_c$ at finite temperature, which may influence the
quantitative feature of the sequential melting \cite{Karsch06},
since non-negligible fraction of $J/\psi$ comes from decay of $\psi'$
and $\chi_c$ in relativistic heavy ion collisions.
This can be done by calculating Wilson coefficients for tensor operators
for these channels.
It should be also noted that
the continuum part of the spectral function may play an important role.
This will be possible by modeling the medium with a gas of
quasi-particle.
One more thing to be done is the extension to higher temperature.
The failure of $T > 1.06T_{\text{c}}$ for $J/\psi$ and $T >
1.04T_{\text{c}}$ for $\eta_c$
originates from the instability of the moment ratio of the OPE side.
The twist-2 gluon condensates becomes larger as temperature increases
and then leads to the breakdown of the stability in the OPE side
including up to dimension 4 (see Fig.~\ref{fig:ope_c}) and
$\mathcal{O}(\alpha_s)$. In Fig.~\ref{fig:ope_cof_a}, we can also see
that the expansion is not good at large $n$ that stabilize the moment
ratio at higher temperature. These facts suggest the necessity of
including higher dimensional operators, which is examined in
Ref.~\cite{Kim01}. However, we do not know a simple way to extract the
temperature dependencies of
the higher dimensional operators from the lattice calculation, as was
done in the present work for dimension 4 operators. The other way of the
extension is to
improve the phenomenological side such that it includes temperature
dependent continuum contribution. The decrease of the scalar gluon
condensates above $T_{\text{c}}$ indicates perturbative contribution
becomes more important at higher temperature. If we can construct a more
appropriate phenomenological side reflecting the nature of the strongly
interacting matter, it will lead to $n$-independent results for physical
parameters until the charmonia really dissolve.

We also note that there are some spaces to improve the analyses for
nuclear matter. Especially, the present analysis shows the mass shift of
$\chi_c$ states are likely to be observed in the forthcoming experiment.
However, the current estimate of the mass shift is not a decisive one;
we have to take the twist-2 contribution into account for a more accurate
estimation.

In summary, we have given a comprehensive analysis on medium-induced
change of the spectral properties of $J/\psi$ and $\eta_c$ in the
hot gluonic medium and the nuclear medium by making use
of QCD sum rules.
In the case of the gluonic medium, our analysis shows
there must be a notable change of mass or width, or both around
$T_{\text{c}}$, caused by the rapid change of the gluon condensates. Although
the present formalism is found to be applicable only up to
$T\simeq 1.06T_{\text{c}}$, the change of mass and width can maximally
reach a few hundred MeV. We have discussed its implication for future
heavy ion experiment at CERN-LHC. As for the nuclear matter case, we
extend the past works to include small but finite width and check the
robustness by varying the scale parameter of the theory. We also examined the
possibility of detecting such mass shifts in the future experiment at
GSI-FAIR. Although $J/\psi$ and $\eta_c$ do not show prominent signals,
$\chi_c$ exhibits more promising results. These analyses give the basis
of future improvements to study the nature of the strongly interacting
matter deeply with charmonia.

\begin{acknowledgments}
 This work was supported by BK21 Program of the Korean Ministry of
 Education. S.~H.~L. was supported by the Korean Research Foundation
 KRF-2006-C00011 and by the Yonsei University research fund. K.~M. would
 like to thank the members of the high energy
 physics group of Waseda University for allowing him to use their
 workstations. We also would like to acknowledge T.~Hatsuda for
 his fruitful comments and discussions.
\end{acknowledgments}

\appendix
\section{Wilson coefficients}
Here we list explicit forms of the Wilson coefficients which appear in
Eq.~\eqref{eq:moment_ope_medium} and are originally given in
Refs.~\cite{Reinders_NPB186} and \cite{Klingl_PRL82}. In the following,
$\rho=\xi/(1+\xi)$ and $F(a,b,c;x)$ is the hypergeometric function $_2
F_1(a,b,c;x)$.

\begin{widetext}

 For the pseudoscalar channel,
 \begin{align}
 A^P_n(\xi) =& \frac{3}{8\pi^2} \frac{2^n (n-1)!}{(2n+1)!!}(4m^2)^{-n}
 (1+\xi)^{-n}F(n,1/2,n+3/2;\rho)\label{eq:w_bare_P}\\
 a^P_n(\xi) =& \frac{(2n+1)!!}{3\cdot 2^{n-1}n!}
 \left[\pi-\frac{1}{2(n+1)}\left(\frac{1}{2}\pi-\frac{3}{4\pi}\right)
 F(n,1,n+2;\rho)\right]
  \frac{1}{F(n,1/2,n+3/2;\rho)}-\left(\frac{1}{2}\pi-\frac{3}{4\pi}\right)
  \nonumber\\
 &+\frac{1}{\pi}\left[\frac{8}{3}-\frac{4}{n}
 \frac{F(n,3/2,n+3/2;\rho)}{F(n,1/2,n+3/2;\rho)}
 -\frac{5}{6}\frac{1}{n+3/2}\frac{F(n,3/2,n+5/2;\rho)}{F(n,1/2,n+3/2;\rho)}
 \right]\nonumber\\
 & -2n\frac{\ln(2+\xi)}{\pi}\frac{(2+\xi)}{(1+\xi)^2}
 \frac{F(n+1,1/2,n+3/2;\rho)}{F(n,1/2,n+3/2;\rho)},\\
 b^P_n(\xi) =& -\frac{n(n+1)(n+2)(n+3)}{2n+3}(1+\xi)^{-1}
 \left[\frac{F(n+1,-3/2,n+5/2;\rho)}{F(n,1/2,n+3/2;\rho)}
 -\frac{6}{n+3}\frac{F(n+1,-1/2,n+5/2;\rho)}{F(n,1/2,n+3/2;\rho)}\right],\\
 c^P_n(\xi) =& b^P_n(\xi)-\frac{4n(n+1)}{(1+\xi)}
 \frac{F(n+1,-1/2,n+3/2;\rho)}{F(n,1/2,n+3/2;\rho)}.
 \end{align}
 Similarly, for the vector channel,
 \begin{align}
 A_n^V(\xi) =& \frac{3}{4\pi^2}\frac{2^n
 (n+1)(n-1)!}{(2n+3)!!}\frac{F(n,1/2,n+5/2;\rho)}{[(4m^2)(1+\xi)]^n},\label{eq:w_bare_V}\\
 a_n^V(\xi) =& \frac{(2n+1)!!}{3\cdot 2^{n-1} n!
 F(n,1/2,n+5/2;\rho)}\left(\frac{2n+3}{2n+2}\right)
 \left[\pi-\left\{\frac{\pi}{3}+\frac{1}{2}
 \left(\frac{\pi}{2}-\frac{3}{4\pi}\right)
 \right\}\frac{F(n,1,n+2;\rho)}{n+1}\right. \nonumber\\
 &\left.+\frac{F(n,2,n+3;\rho)}{3(n+1)(n+2)}\left(\frac{\pi}{2}-\frac{3}{4\pi}\right)\right]
 -\left(\frac{\pi}{2}-\frac{3}{4\pi}\right)
 -2n\frac{\ln(2+\xi)}{\pi}\frac{(2+\xi)}{(1+\xi)^2}\frac{F(n+1,1/2,n+7/2;\rho)}{F(n,1/2,n+5/2;\rho)},
 \\
 b_n^V(\xi) =&
 -\frac{n(n+1)(n+2)(n+3)}{(2n+5)(1+\xi)^2}\frac{F(n+2,-1/2,n+7/2;\rho)}{F(n,1/2,n+5/2;\rho)},\\
 c_n^V(\xi) =& b_n^V(\xi)-\frac{4n(n+1)}{3(2n+5)(1+\xi)^2}\frac{F(n+2,3/2,n+7/2;\rho)}{F(n,1/2,n+5/2;\rho)}.
 \end{align}
 In Eqs.~\eqref{eq:w_bare_P} and \eqref{eq:w_bare_V}, $m$ is the running quark
 mass $m=m_c(p^2 = -(\xi+1)m_c^2)$ which is given by \cite{Novikov78},
 \begin{equation}
 \frac{m_c(\xi)}{m_c(\xi=0)} =
  1-\frac{\alpha_{\text{s}}}{\pi}\left[\frac{2+\xi}{1+\xi}\ln(2+\xi)-2\ln
				  2\right]
 \end{equation}
\end{widetext}

\end{document}